%% file: NGdraft3.tex
\documentclass[prd,a4paper,eqsecnum,nofootinbib,preprint,floatfix]{revtex4}
\usepackage{epsfig}
\usepackage{colordvi}
\usepackage{graphicx}
\usepackage{bm}
\usepackage{multirow}
\reversemarginpar

\def\<{\langle}
\def\>{\rangle}

\newcommand{\text}{\rm}
\def\Tr{{\rm Tr}\,}

\def\Eq#1{Eq.~(\ref{#1})}

\begin{document}

\vspace*{0.5in}

\title{The closed string spectrum of \\ $SU(N)$ gauge theories in $2+1$ dimensions \vspace*{0.25in}}

\author{Andreas Athenodorou, Barak Bringoltz and Michael Teper\textsc{}\\
\vspace*{0.25in}}

\affiliation{Rudolf Peierls Centre for Theoretical Physics,
University  of  Oxford,\\
1 Keble Road, Oxford, OX1 3NP, UK 
\vspace*{0.6in}
}

\begin{abstract}
We use lattice techniques to study the closed-string spectrum of
$SU(N)$ gauge theories in $2+1$ dimensions. We calculate the
energies of the lowest lying $\sim 30$  states for strings with lengths between $l\sim 0.45$ fm and $l\sim 3$ fm, and compare to different theoretical predictions. We obtain unambiguous evidence that the closed-strings are in the universality class of the Nambu-Goto free bosonic string. Moreover, we clearly see that our data can be described by a covariant string theory with a small/moderate correction down to very short distance scales, and possibly on all distance scales at large-$N$.

\end{abstract}
\maketitle

\section{Introduction}
\label{intro}

The generation of electric flux tubes in confining gauge theories
is a basic phenomena that characterises the vacuum
of these theories. In this paper we study the energy spectrum of these
flux tubes
using lattice techniques. We are mainly motivated by the following two
reasons.

Firstly, a flux-tube whose length $l$ is much larger than its
width, is expected to be described by an effective low energy string action
$S_{\rm eff}$. Establishing what is the structure of $S_{\rm eff}$ is of
fundamental interest, and can be done by studying the energy spectrum
of the flux-tube.
More precisely, while the energy levels $E_n$ of a string with tension $\sigma$ obey
\begin{equation}
E_n(l) \stackrel{l\to \infty}{=} \sigma \, l, \label{sigma_def}
\end{equation}
for finite $l$ there
are corrections to \Eq{sigma_def} that reflect various properties of $S_{\rm
  eff}$. In particular, the universal coefficient of the $O(1/l)$
`Luscher-term' is determined by the number of the massless degrees
of freedom propagating on the worldsheet of the string, thus
revealing the IR universality class of $S_{\rm eff}$
\cite{LW_old}. In contrast, other details
 of the spectrum, like its degeneracies and the form of the
 subleading contributions to the Luscher term, are in general not universal
and depend on the particulars of the non-renormalizable terms in
$S_{\rm eff}$. Consequently, by studying the $l$-dependence of the string spectrum, one can directly learn about $S_{\rm
  eff}$, and about the length $l_{\rm string}$ above which $S_{\rm
  eff}$ begins to be a good description and a string is formed.

Our interest in the flux-tube spectrum is also driven by the
following more practical reason. An essential step in any lattice
study is to calculate the lattice spacing $a$ in physical units.
This is sometimes done by
 extracting the dimensionless combination $a \sqrt{\sigma}$ from a flux-tube's ground state energy given in lattice units
 $a E_{n=0}$. Consequently, this requires that we know how $E_0$ depends on $\sigma$
 and $l$. To date, this dependence was approximated by correcting
 \Eq{sigma_def} with {\em just} the Luscher term. For the purpose of high precision
lattice studies, however, this approximation can become insufficient and
 it is imperative to know what are the subleading corrections to \Eq{sigma_def} that go beyond the Luscher term.

In this work we focus on the spectrum of {\em closed} flux-tubes
in pure $SU(N)$ gauge theories in $D=2+1$ dimensions. The
flux-tubes that we study have lengths that range from
$l\simeq 0.45$fm to $l\simeq 3$fm, while the lattices we use have
spacings that range from $\sim 0.06$fm to $\sim 0.2$fm, depending
on the value of $N$ and $l$. (Here, despite working in $2+1$ dimensions and in pure gauge theories, we choose to {\em define} $1$fm through the convention $\sigma
\equiv (440 \,{\rm MeV})^2$.)

The gauge groups that we study have $N=2,3,4,5,6,8$, and here we
are motivated by the special role that the large-$N$ limit plays
in the physics of confinement; From a field theory point of view,
several processes that cannot be described by a simple low energy
effective string theory, such as glueball-string mixing,
string/anti-string mixing and deconfinement/instability of short
strings, become less important
 with increasing $N$ (see the discussion in Section~\ref{short-distance}). This
is expected to make the description of the flux-tube in terms of
$S_{\rm eff}$ better and simplify the form of the latter. From a
more general string/gauge duality point of view (for example see
\cite{PolchinskiQCD,Aharony}), it is the large-$N$ limit of QCD
that one may hope to describe by a string theory. This makes
non-perturbative information on this limit important both to guide
the searches for such a dual string theory, as well as to
understand it beyond the supergravity approximation.

There are two stages to our calculation which henceforth we refer
to as A and B. In stage A we perform high precision
measurements of the energy of the flux-tube's ground state. In
particular, we aim to isolate the different corrections to
\Eq{sigma_def} and, while controlling them, extract a value for
$\sigma$. We then use the latter to compare our data to different
theoretical predictions, and especially to the Nambu-Goto (NG) free string model. This stage of our work has two immediate practical
implications. Firstly, it allows us to test the analytic work in
\cite{LW_new} which predicts that in $D=2+1$ the $1/l^3$
correction to $E_n$ has a universal coefficient, being precisely that predicted by the NG model.\footnote{Note
that, using the method of \cite{PS_old}, the authors of
\cite{PS_new} assert that this universality persists for a general
value of $D$.} Secondly, a precise control of the corrections to
\Eq{sigma_def} has enabled us to perform, in a companion study \cite{KN},
 a precise test of the Karabali-Kim-Nair prediction for the value of
$\sigma$ in $2+1$ dimensions \cite{KKN}.

 In stage B we put our emphasis on the excites states in the
 spectrum and measure the energy of the lowest $\sim 30$
 states.
This allows to extend the comparison with theoretical
 predictions to states with more quantum numbers and a nontrivial
degeneracy structure.

The study of confining flux-tubes with lattice techniques has been an active field of research for the past three decades, and we refer the reader to some recent papers \cite{strings-lattice-review}. These include works that vary in the selection of the gauge group, the number of space-time dimensions, and the boundary conditions imposed on the flux-tube.  Other papers which are particularly relevant to our current study are mentioned later on in the text.

The following is the outline of this letter. We begin in
Section~\ref{method} by describing our methodology, proceed to
Section~\ref{theory} where we discuss the theoretical expectations
for the string spectrum, and move to
Sections~\ref{gs}-\ref{spectrum} where we present the results of the
calculations. In Section~\ref{conclusions} we
summarise our results and make a few remarks on their theoretical
and practical implications.

\section{Methodology}
\label{method}

In this section we describe a few aspects of our methodology. We
begin with the lattice construction, proceed to discuss our
general strategy to extract the energy spectrum from correlation
functions, and end by listing the main systematic
errors and by expanding on how we control them.

\subsection{Lattice construction}
\label{lattice}

We define the gauge theory on a discretized periodic Euclidean
three dimensional space-time lattice, with spacing $a$. The fields are $SU(N)$ matrices assigned to the links of the lattice, and the Euclidean
path integral is given by
\begin{equation}
Z=\int DU \exp{\left( -\beta S_{\rm W}\right)}.
\label{Z}
\end{equation}
Here $\beta$ is the dimensionless lattice coupling, and for our action is related to
the dimensionful coupling $g^2$ by
\begin{equation}
\lim_{a\to 0}\beta=\frac{2N}{ag^2}.
\label{betag}
\end{equation}
In the large--$N$ limit, the 't Hooft coupling $\lambda=g^2N$
is kept fixed, and so we must scale $\beta=2N^2/\lambda \propto N^2$
in order to keep the lattice spacing fixed (up to $O(1/N^2)$
corrections). The action we choose to use is the standard Wilson action
\begin{equation}
S_{\rm W}=\sum_P \left[ 1- \frac1N {\rm Re}\Tr{U_P} \right],
\label{eq:SW}
\end{equation}
where $P$ is a lattice plaquette index, and $U_P$ is the plaquette
variable obtained by multiplying link variables along the circumference
of a fundamental plaquette. We calculate observables by performing
Monte-Carlo simulations of \Eq{Z}, in which we use Cabibbo-Marinari updates of the link matrices with a mixture of
Kennedy-Pendelton heat bath and over-relaxation steps for
all the $SU(2)$ subgroups of $SU(N)$.

In stage A we have studied $N=2,3,4,5,6,8$ with three lattice
spacings, $a\simeq 0.06,0.08,0.11$ fm. In stage B we studied
$SU(3)$ with $a\simeq 0.04,0.08$ fm, and $SU(6)$ with $a\simeq
0.08$ fm. The string lengths $l$ in both stages ranged between
$\sim 0.45$ fm and $\sim 3$ fm, depending on the values of $N$ and
$a$. For more details on the lattice parameters of our field
configurations see Tables~\ref{fits-gs-partA}-\ref{fits-gs-partB}.

\subsection{General strategy}
\label{strategy}

Since we are mainly interested in the way the flux-tube spectrum
reflects the properties of a standard low energy string theory of the NG type,
we restrict ourselves to the spectrum of {\em closed} flux-tubes. By this
restriction we avoid a class of short-distance contributions to the
energies, such as the Coulomb interaction between sources, that cannot be easily accommodated in an effective string theory
(see the discussion in
Section~\ref{short-distance}).

We calculate the energies of flux tubes that are closed around a spatial torus. We do so from the correlators of suitably smeared Polyakov loops that wind around
one of the spatial tori and that have vanishing transverse
momentum. This is a standard technique
\cite{Teper_Nd3,Teper_Lucini} with the smearing/blocking designed
to enhance the projection of our operators onto the physical
states. (We use a scheme that is the obvious dimensional reduction
of the one in \cite{LTW_ops}.) We classify our operators using the
following quantum numbers
\begin{enumerate}
\item $P=\pm$ : Parity in the direction transverse to Euclidean time and to
  the string contour.
\item $q = 0,\pm 1, \pm 2,\dots$ : The momentum, in units of $2\pi/l$,
  along the string.
  \end{enumerate}
For each combination of these quanta we construct the full
correlation matrix over our space of loop operators, and use it to obtain best estimates for the
string states using a variational method
applied to the transfer matrix $\hat{T}=e^{-aH}$ -- again a
standard technique \cite{var,Teper_Nd3,Teper_Lucini,LTW_ops}.

The size $N_{o}$ of our $N_{o}\times N_{o}$ correlation matrices depend on the states we
are interested in. In stage A, where we focus on the lowest state,
our correlations are built from Polyakov loops that wind once
around the torus in straight paths and in all possible blocking
levels. Consequently, this means that we restrict ourselves to states with
 $q=0$ and $P=+$, and this provides us with $N_{o}=3-5$ states. In
 stage B we include Polyakov loops with many different paths (wave-like
 paths, pulse-like paths, etc.). This increases our number of
operators to $N_{o}\simeq 80-200$ and allows us to probe states with
negative $P$ and nonzero $q$. Finally, an extension of our
calculation for the ground state to $w>1$ will be reported in
\cite{BT-kstrings}.

Once we obtain the string energies for different values of $l$, we
fit the ground state energy in powers of $1/l$ and obtain an
 estimate for the function $E_0(l,\sigma)$. With this empirical formula we extract the string tension in lattice units, $a^2 \sigma$, for
each data set. Substituting the results in different theoretical
predictions for the spectrum, we conclude by comparing the latter
with our data.

\subsection{Systematic errors}
\label{systematics}

We control several systematic errors in different stages of our
study. These include the way we extract the string energies from
correlation functions, the way our results approach the
infinite volume and continuum limits, and the way one may interpret the
results for {\em small} values of
$l$ as reflecting an effective string picture.
In the next three sub-sections we expand on each systematic error,
and on the way that we control them.

\subsubsection{Extracting string energies from correlation functions}
\label{contamination}

The output of the variational technique is a set of
operators that couples ``best'' to a set of low energy states.
For example, our best operator for the string's ground state has
typically an
overlap $\sim 99\%$ onto that state so that the normalised `ground
state' correlation function satisfies
\begin{equation}
C(t) = (1-|\epsilon|) \exp\{-E_0(l)t\}
+ |\epsilon_1| \exp\{-E_1(l)t\} + ...
\quad ; \quad
\sum_{i} |\epsilon_i| = |\epsilon| \sim 0.01
\label{correl}
\end{equation}
where $E_0,E_1$ are the ground and first excited state string
energies. (Since our time-torus is finite, we use cosh fits rather
than simple exponentials, although in practice we use $L_t$ large
enough for any contributions around the `back' of the torus to be
negligible.) To extract $E_0$ from this correlator one can fit
with a single exponential for $t\geq t_0$, discarding $t<t_0$, and choosing $t_0$ to be the minimum value so that a statistically acceptable fit is obtained.
This is a reasonable approach and one followed in
\cite{Teper_Nd3,Teper_Lucini}. However it neglects the systematic
error arising from the fact that there is certainly some excited
state contribution as demonstrated, for example, by the fact that
one cannot obtain a good fit with a single exponential from $t=0$.
To control this systematic error we also perform fits with two
exponentials, with a fixed mass $E^*$ for the excited state,
resulting in a mass $E_0(E^*)$ for the ground state. Typically
$E_0(E^*)$ is smallest when $E^*$ is as small as possible, i.e.
$E^* = E_1$, and is largest when  $E^*=\infty$, i.e. effectively a
single-exponential fit. So the true value typically satisfies:
\begin{equation}
E_0(E_1) \le E^{\rm true}_0 \le E_0(\infty).
\end{equation}
From here on, we refer to the single-cosh fitting procedure by
`S', and to the double-cosh fitting procedure, by `D'. This particular systematic error becomes more important as the string length $l$ increases because the overlap of our lattice operators typically decreases with increasing $l$, and the excited state energy approaches the ground state energy. In practice, we find that as long as $l\stackrel{<}{_\sim}5/\sqrt{\sigma}$ one can neglect this systematic error at the level of our current statistical errors. Consequently we present results from the `D' fitting procedure only when $l>5/\sqrt{\sigma}$.

\subsubsection{Finite volume and discretisation effects}
\label{finite-V}

To avoid finite volume effects in the calculation
of the closed string spectrum we follow \cite{TM04} and increase the
transverse and temporal extents of the torus, when we decreases
the length of the string $l$. We performed explicit finite
volume checks for a restricted set of parameters, and found that the transverse volumes used in our study are large enough to avoid any observable effects at the level of our statistical accuracy. To check for finite lattice spacing effects
we perform several of our calculations with different lattice
spacings.

\subsubsection{A string interpretation of the flux-tube spectrum for
  short lengths  ? }
\label{short-distance}

The energy spectrum of confining flux-tube is expected to be much
more complicated than that of a simple effective string and to approach
the latter only at large $l$. Therefore, in an ideal calculation
one would study the $1/l$ terms in the string energies only for
$l\gg 1/\sqrt{\sigma}$. In practice, however, the $1/l$ terms are
numerically small and a useful measurement at large $l$ would
require an unrealistically large statistical sample. Instead, we
study strings with $1/\sqrt{\sigma} \stackrel{<}{_\sim} l
\stackrel{<}{_\sim} 6/\sqrt{\sigma}$ for which the corrections are
not negligible and can be reasonably fitted. This procedure is not
free of ambiguities since the short-$l$ spectrum may be sensitive
to phenomena not accommodated in an effective string theory, but
as we explain below, it is unlikely that this ambiguity is
significant in our calculation.

The first obvious source of short-distance ``contamination'' in
studies of flux-tube spectra is the presence of a Coulomb
interaction between the static charges at the ends of {\em open}
flux-tubes. This can be
a significant portion of the total energy when the flux tube is
short, so it is not clear whether the deviations from the infinite
length limit seen in the open channel (for example see
\cite{Meyer-forces} and reference within) are due to this Coulomb
interaction or whether they reflect string interactions. Our
calculation, however, is free of this contamination simply because
we study {\em closed} flux-tubes and by construction these  do not
have static charges and a Coulomb interaction.

A different type of non-stringy phenomenon that does occur in the
closed channel is the deconfinement transition. In $2+1$
dimensions this finite temperature transition takes place at a
temperature $T=T_d\simeq 0.9/\sqrt{\sigma}$
\cite{LiddleTeper_Tcd3}. By identifying the length of the compact
direction, $l$, with the inverse temperature $T^{-1}$, this means
that our calculation necessarily breaks down when $l<1/T_d\simeq
1.1 1/\sqrt{\sigma}$, but also that interpreting the flux-tube
energies as coming from the dynamics of $S_{\rm eff}$ may be
questionable in the confined phase, when $l\simeq (1/T_d)^+$. This
is particularly true for $N=2,3$ when the transition is second
order and the $l\to \left(1/T_d\right)^+$ behaviour of $E_0(l)$ will be governed by appropriate critical exponents, and may also occur for $N=4,5$, where the first order transition is relatively weak. In contrast, for $N\ge 6$, the transition is {\rm strongly} first order and it is quite possible that a
confining string description exists even when $l\simeq
(1/T_d)^+$.\footnote{As an extreme example, note that for very
large values of $N$ and in
  $3+1$ dimensions, it is possible to study confining flux-tubes even {\em
    below} $1/T_d$ \cite{BT-Hagedorn}.}

Another short-distance phenomenon involves glueballs : by
decreasing $l$ the energy needed to excite the string grows (see
Section~\ref{theory}) and if we take the spectrum of the free
bosonic string prediction as a guide, then the threshold for the
first excited state to emit the lightest $0^{++}$ glueball and
decay to the ground state is reached when $l\simeq
1.53/\sqrt{\sigma}$. In practice, however, since the amplitudes of these
mixing processes are subleading in $1/N^2$, they may be suppressed
even for $SU(3)$.

To conclude, the short-distance non-stringy phenomena that
may contaminate the string picture interpretation of the closed flux-tube
spectrum, go away at large-$N$. Hence, by performing our
calculations for increasingly large values of $N$ we explicitly
check how large are these effects and thereby control them.

\section{Theoretical expectations for $E_n(L)$}
\label{theory}

Let us now discuss the theoretical predictions to which we compare the
measured flux-tube spectrum.
\subsection{The spectrum of the Nambu-Goto model}
\label{section_NG}
The action of the NG model \cite{NGpapers}
is the area of the worldsheet swept by the propagation of the
string.
 Due to the Weyl anomaly this model is
quantum-mechanically consistent only in the critical dimension
$D=26$ (see for example \cite{Polchinski}), but since this anomaly
is suppressed for long strings \cite{Olesen} it can still be
considered as an effective low energy model.

The single string states can be characterized by the number of
times $w$ that the string winds around the torus. The spectrum in
each case corresponds to the transverse oscillations of the
worldsheet that correspond to movers that travel clockwise and anti-clockwise along the
string. Thus the string states are characterised by $w$, by the
occupation number $n_{L(R)}(k)$ of left(right) movers that carry energy $k$, and also by the centre of mass momentum $\vec p_{\rm c.m.}$. By projecting to zero transverse momentum $\left(\vec p_{\rm c.m.}\right)_\perp$ we are left only with
the momentum along the string axis which is quantized in units of
$2\pi q/l$ with $q=0,\pm 1,\pm 2,\dots$ for a string of length $l$. These quanta are not
independent of $n_{L,R}$ and obey the level matching
constraint\footnote{This condition constraints the physical states
to be invariant under the gauged diff-invariance of the action \cite{Polchinski}, and is effectively momentum conservation.}
\begin{eqnarray}
N_L- N_R &=& q w,\label{levels}
\end{eqnarray}
where $N_{L(R)}$ enumerates the momentum contribution of the
left(right) movers in a certain state as follows
\begin{equation}
N_L= \sum_{k>0} \,\,\sum_{n_L(k)>0} n_L(k) \,k, \qquad
N_R= \sum_{k'>0} \,\,\sum_{n_R(k')>0} n_R(k') \,k'.
\end{equation}

It is customary to characterise the string states as
irreducible representations of the $SO(D-2)$ symmetry that rotates
the spatial directions transverse to the string axis. In our
$D=2+1$ dimensional case, this group becomes the transverse parity
$P$ and acts by assigning a minus sign for each mover on the
worldsheet. As a result the string states are eigenvectors of $P$
with eigenvalues
\begin{equation}
P = \left(-1\right)^{\sum_{i=1}
  n_L(k_i) + \sum_{j=1} n_R(k'_j)} .
\end{equation}

Finally, the energy of a closed-string state with the above quanta is (here
we write it for a general number of spacetime dimensions $D$)
\begin{eqnarray}
\left(E_{N_L,N_R,q,w}\right)^2 &=& \left(\sigma \,l w\right)^2 + 8 \pi \sigma
\left( \frac{N_L+N_R}2 - \frac{D-2}{24}\right) + \left(\frac{2\pi
q}{l}\right)^2.\label{NG0}
\end{eqnarray}

\subsection{Effective string theories}
\label{section_Seff}
Since in $2+1$ dimensions the NG string is at best an
effective low-energy string theory, it makes sense to generalise it
and write the most general form of an effective string action
$S_{\rm eff}$ consistent with the symmetries of the flux-tube
system. This was done some time ago for the $w=1$ and $q=0$ in
\cite{LW_old} and the spectrum obtained for a general number of
space-time dimensions $D$ was
\begin{equation}
E_{n} = \sigma l + \frac{4\pi}{l} \left(n-\frac{D-2}{24}\right) + O(1/l^2),\label{luscher0}
\end{equation}
with $n=0,1,2,\dots$. Here the second term on the right hand side
of \Eq{luscher0} is known as the Luscher term and is expected to
be universal and independent of the particulars of IR-irrelevant
interactions of the low energy effective string theories. Indeed,
it can be easily verified that the NG model
obeys this universality by expanding the square-root of \Eq{NG0}
to leading order in $1/l$.

The work \cite{LW_old} was more recently extended in
\cite{LW_new}, where the authors established and used a certain
open-closed string duality of the effective string theory. Using
this duality they showed that for any number of spacetime
dimensions the $O(1/l^2)$ is
 absent from \Eq{luscher0}, and that in $D=2+1$ the $O(1/l^3)$ has a
 universal coefficient. Consequently, in $2+1$ dimensions,
 \Eq{luscher0} is extended to
\begin{equation}
E_{n} = \sigma l + \frac{4\pi}{l} \left(n-\frac{1}{24}\right) - \frac{8\pi^2}{\sigma \,l^3} \left(n-\frac{1}{24}\right)^2 + O(1/l^4).\label{luscher1}
\end{equation}
In the context of the covariant string description such as that of Section~\ref{section_NG}, \Eq{luscher1} implies that 
\begin{equation}
E_{n} = \sqrt{\left(\sigma l\right)^2 + 8\pi\sigma \left(n-\frac{1}{24}\right) + O(1/l^3)},\label{luscher2}
\end{equation}
 which is a particularly convenient form since the two
first terms under the square root are the prediction of the NG
model for the energy squared, that we find to be a very good approximation (see below).
A different approach
that also leads to similar conclusions is the Polchinski-Strominger
 effective string theory \cite{PS_old}. While it originally yielded
 \Eq{luscher0}, it was used recently in \cite{PS_new} to extend the
 analysis to higher powers of $1/l$ and leads to the same conclusions
 as \cite{LW_new}, but for all values of $D$.

Motivated by these recent
developments, we take our main fitting ansatz for the spectrum to be
\begin{equation}
E_{\rm fit} = \sqrt{E^2_{NG} - \sigma
\frac{C_p}{\left(l\sqrt{\sigma}\right)^{p}}}\quad ; \quad p\ge 3,\label{fit}
\end{equation}
where $E^2_{NG}$ is the NG prediction given by
\Eq{NG0} and where $C_p$ are dimensionless coefficients that in
general can depend on the quantum numbers of the state.

Let us pause and make the following comment about the relation between \Eq{luscher1} and \Eq{luscher2}. Consider performing a large-$l$ expansion of the square-root in \Eq{luscher2}. The result would include not only \Eq{luscher1}, but also many additional terms that begin at $O(1/l^5)$,  and that come from the expansion of the second term under the square-root. To see whether these extra terms are important, we can use our data and compare its deviations from \Eq{luscher1} to its deviations from \Eq{luscher2}. If we find that the latter are systematically smaller, this would then reflect the naturalness of adding a correction term in a covariant way (correcting $E^2_n$ as in \Eq{luscher2} and \Eq{fit}) rather than correcting the energy $E_n$ itself (as \Eq{luscher1} does). From a field-theoretical point of view, such as the one in \cite{LW_new}, it seems that it would be hard to get a prediction of the form of \Eq{luscher2} since it can be viewed as an implicit resummation of an infinite series of powers of the Luscher term.

\section{A precise measurement of the ground state energy $E_0(\sigma,l)$}
\label{gs}

\begin{figure}[htb]
\centerline{
\includegraphics[width=9cm]{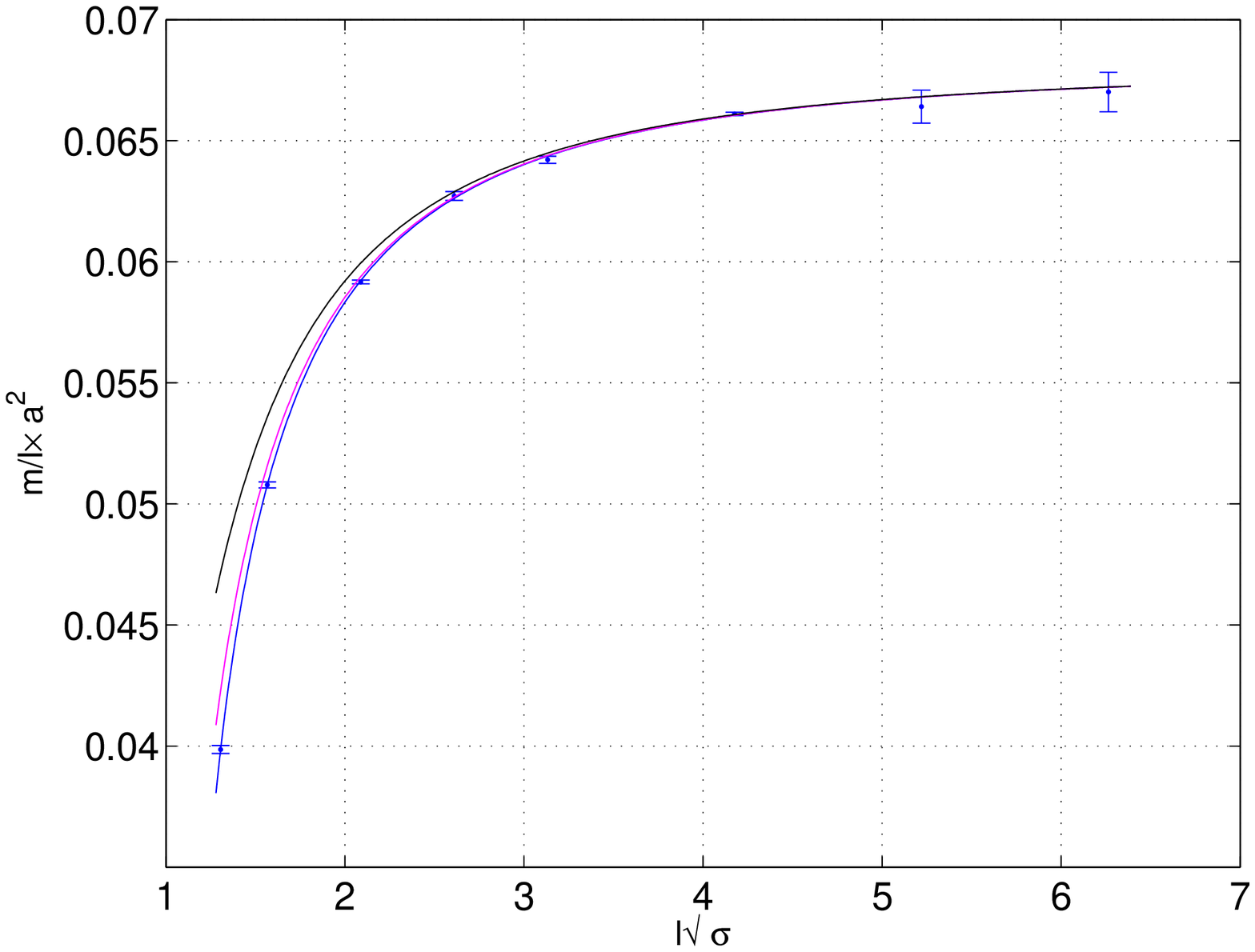} \quad
\includegraphics[width=9cm,height=7.075cm]{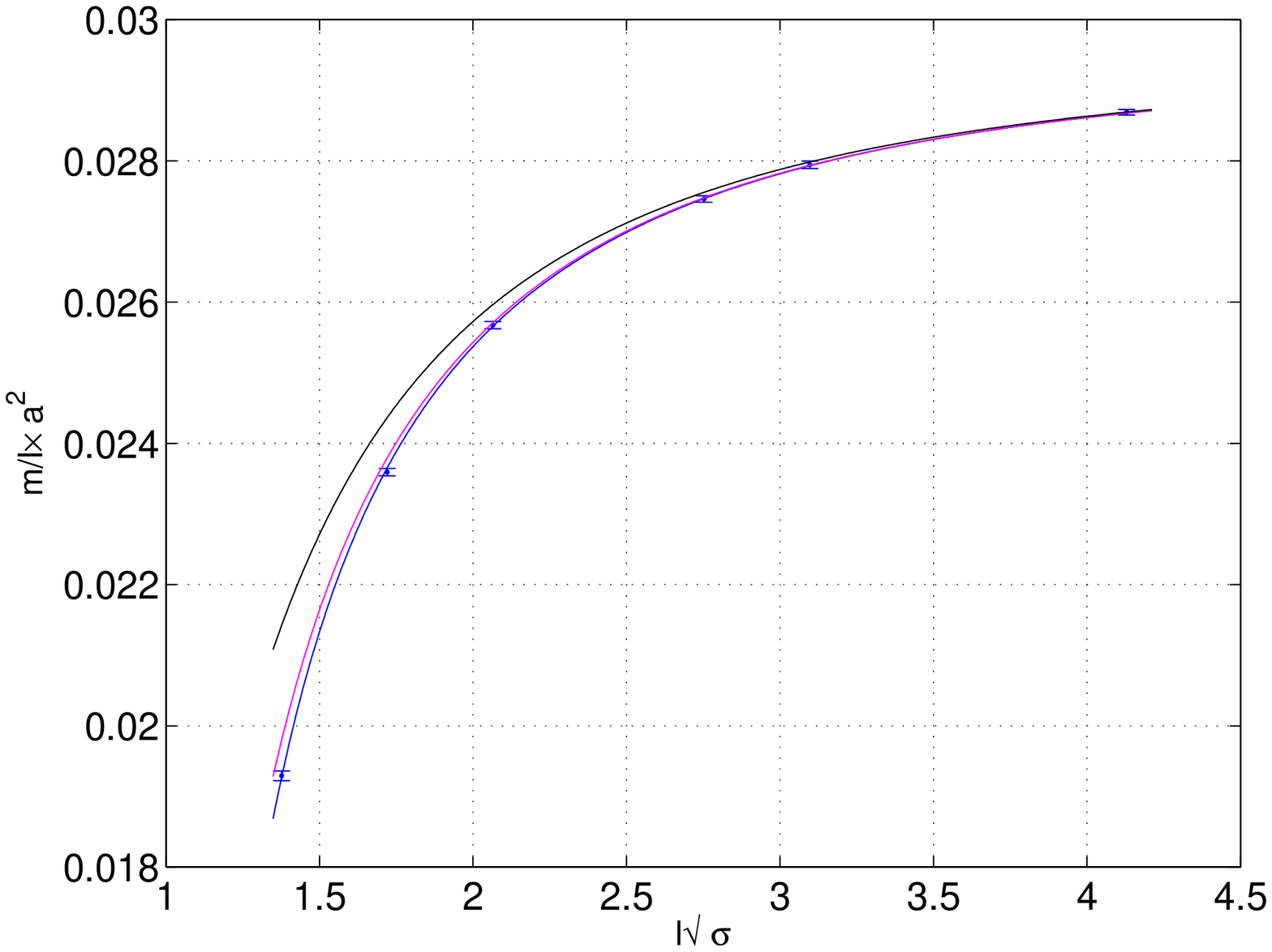}
}
\caption{The ground state energy per unit length, in lattice units,
   vs. the string length for $SU(3)$ and $\beta=14.7172$ (left panel)
   and for $SU(6)$ and $\beta=90.00$ (right panel). All the data points are the result of the single exponential fits, except when $l>5/\sqrt{\sigma}$, where we use the results from double exponential fits (see
  Section~\ref{contamination}).
The blue lines are the results of using the fitting ansatz
\Eq{fit}. Using the values we obtain for the string tensions we
plot the NG predictions of \Eq{NG0}  (magenta lines) and of the
Luscher formula in \Eq{luscher0} (black lines).}
 \label{gsfig}
\end{figure}

We performed high precision calculations of the dependence of the
ground state energy on the length of the string. The calculations 
were done for the numbers of colors $N=2,3,4,5,6,8$. The lengths
of the strings were restricted to obey $l>1/T_d$, with the
deconfinement transition temperature $T_d$ given in
\cite{LiddleTeper_Tcd3}. As examples, we present the results for
$N=3$ and $N=6$ with $\beta=14.7172$ and $\beta=90.00$
respectively, in Fig.~\ref{gsfig}.

Fitting the data with \Eq{fit} and $p=3$ we obtain acceptable fits only when we use the `D' data for strings with $l\stackrel{>}{_\sim}5/\sqrt{\sigma}$. This indicates that for long
strings, for which the first excited energy is relatively close to
the ground state energy, the control over contamination of the
excited states is crucial. The results for the other gauge groups
and lattice spacings are similar to the $SU(6)$ results, and we
present the parameters and goodness of fits of the form in \Eq{fit} with $p=3$
 in Table~\ref{fits-gs-partA}.
  
From the table one can also see that the
$N$-dependence and the lattice spacing dependence of the correction $C_3$ is
moderate/small. We have also performed fits with $p=4$, and obtained comparable $\chi^2$, which means that with the
current data for the ground state, we cannot unambiguously determine the power of the correction term in \Eq{fit}.
\begin{table}[htb]
\centering{
\begin{tabular}{|c|c|c|c|} \hline \hline
Configuration details       &  $a^2\sigma$ & $C_3$ &
Confidence level \\ \hline \hline
$SU(2),\,\beta=5.6$  & $0.074636(25)$ & $0.0100(41)$ &  $53\%$ \\  \hline
$SU(3),\,\beta=14.7172$  & $0.068127(47)$ & $0.1633(112)$ &  $87\%$ \\  \hline
$SU(4),\,\beta=28.00$ & $0.063346(73)$ & $0.2943(282)$ &  $99\%$ \\ \hline
$SU(4),\,\beta=50.00$ & $0.017184(14)$ & \, $0.1886(338)$ \, &  $10\%$ \\ \hline
$SU(5),\,\beta=80.00$ & $0.016874(12)$ & $0.0554(139)$ &  $68\%$ \\  \hline
$SU(6),\,\beta=59.40$ & $0.077460(81)$ & $0.1098(169)$ &  $16\%$ \\ \hline
$SU(6),\,\beta=90.00$ & $0.029601(23)$ & $0.1163(159)$ &  $88\%$ \\ \hline
$SU(8),\,\beta=108.00$ & \, $0.075351(141)$ \, & $0.1865(575)$ &  $68\%$ \\ \hline
$SU(8),\,\beta=192.00$ & $0.020200(24)$ & $0.0864(109)$ &  --  \\ \hline
\end{tabular}
} \caption{The lattice parameters $\beta$ and $a^2\sigma$ and
$C_3$ in the fit \Eq{fit}. The fits were performed for the $S$ data sets except for strings of length $l>5/\sqrt{\sigma}$, where we used the $D$ type of data (Single and Double exponential fitting of the correlation functions - see Section~\ref{contamination}).  For $SU(8)$ and
  $\beta=192.00$ our confidence level is small, and comes from a large scatter of the data points around the fit. This indicates that our errors are
  probably underestimated in this case.}
\label{fits-gs-partA}
\end{table}

As Fig.~\ref{gsfig} demonstrates, the Luscher
term formula and the NG formula are
indistinguishable from each other and agree with the fit at large
values of $l$. The situation at small values of $l$, however, is
completely different. There, the Luscher term is clearly
insufficient to describe the data, indicating the importance of
the subleading corrections to \Eq{luscher0} for
$l\stackrel{<}{_\sim}3/\sqrt{\sigma}\simeq 1.35$ fm. In contrast, the NG
prediction is very good, even at the lowest value of $l$. This
remarkable fact also explains why we choose \Eq{fit} to be our
fitting ansatz, rather than using a fit which assumes that the
energy itself is a power series in $1/l$.

Finally, we perform the following two type of fits
\begin{eqnarray}
{\rm Fit \, 1.} &\qquad & E_0(l,\sigma) = \sigma l -
\frac{\pi}{6}\times C^{(1)}_{\rm eff},\label{Ceff1fit}\\
{\rm Fit \, 2.} &\qquad & E_0(l,\sigma) = \sqrt{\left(\sigma l\right)^2 -
\frac{\pi\sigma}{3}\times C^{(2)}_{\rm eff}},\label{Ceff2fit}
\end{eqnarray}
to pairs of adjacent values of $l$ in plots of the type of Fig.~\ref{gsfig}.

\begin{figure}[htb]
\centerline{
\includegraphics[width=9cm]{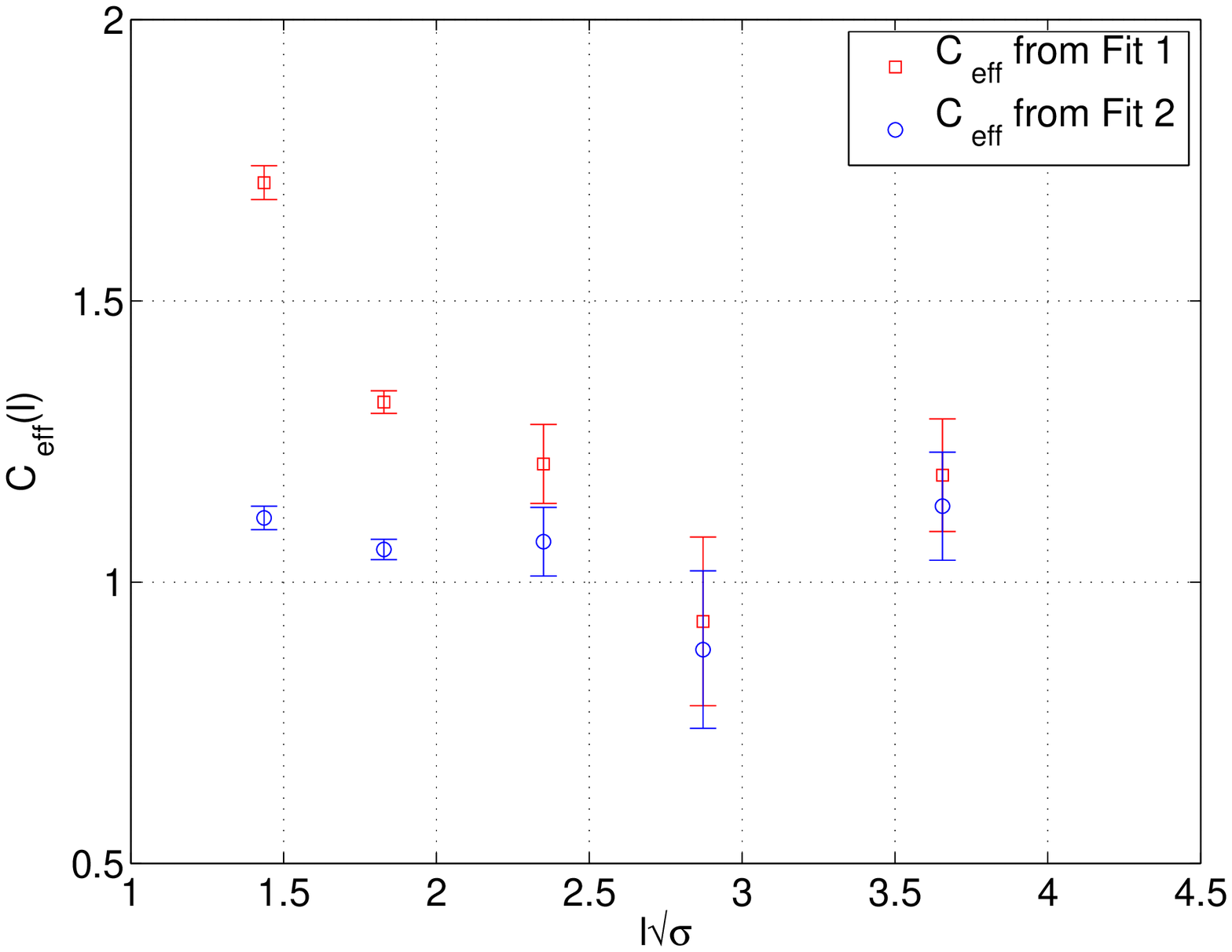} 
\quad \includegraphics[width=9cm]{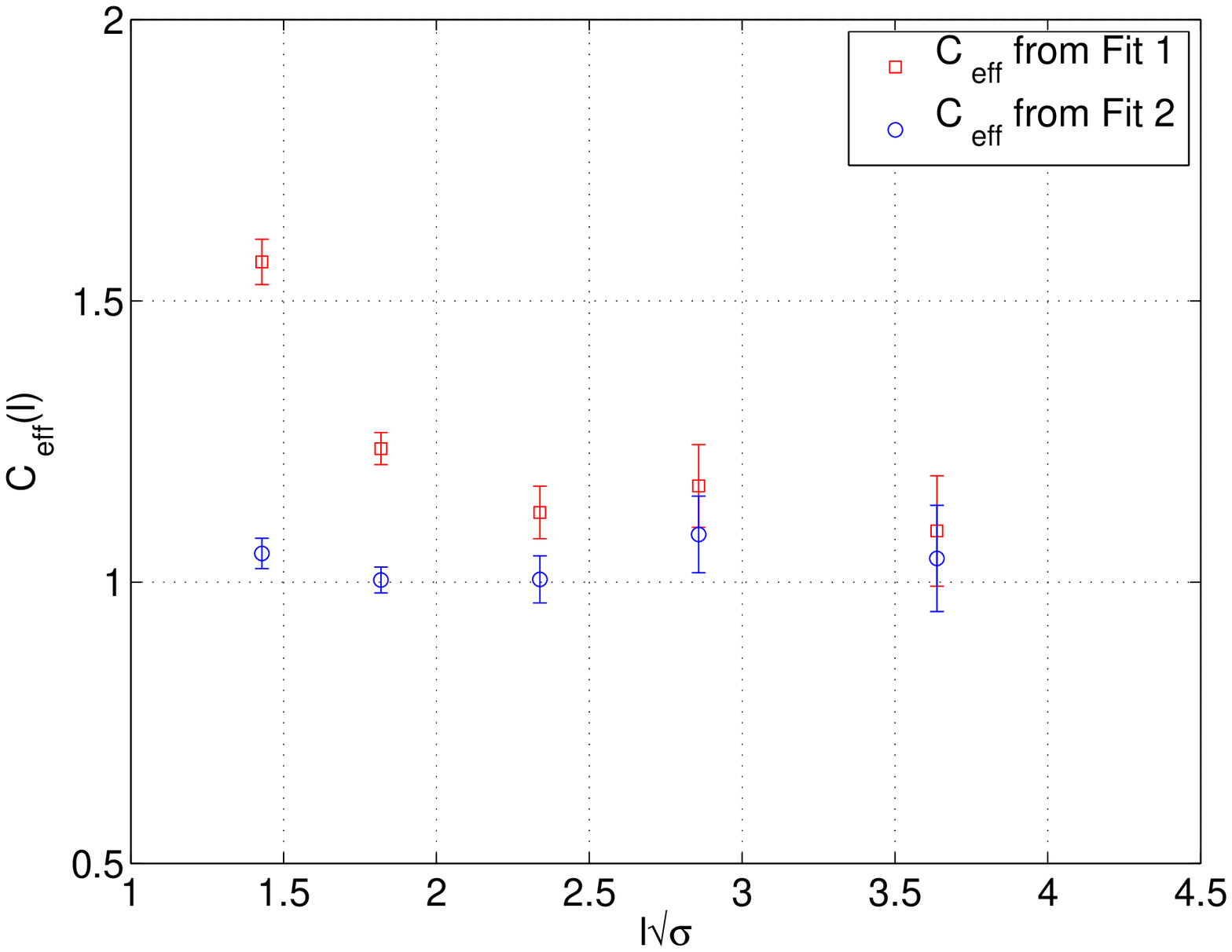}
}
\caption{The effective central charges $C^{(1,2)}_{\rm eff}$ vs. the
  string length for the 'S' data points, for $SU(3)$ and
  $\beta=14.7172$ in the left panel, and for $SU(5)$ and
  $\beta=80.00$ in the right panel.}
\label{Cefffig}
\end{figure}

The effective central charge $C_{\rm eff}(l)$ in both type of fits
is expected to approach a value of unity for large values of $l$,
but to deviate from $1$ when $l$ decreases because of the higher order $O(1/l^3)$ terms. The fit in
\Eq{Ceff1fit} is the one usually performed (see for example
\cite{Ceff}) and is the analog of the effective central charge fits
performed in the open channel \cite{LW_new,Meyer-forces,Majumdar}.
 In contrast, the ansatz
in \Eq{Ceff2fit} assumes that most of $C^{(1)}_{\rm eff}$'s
deviation from $1$ comes from the $1/l$ terms that are
accommodated in the NG prediction. Indeed, this assumption is
confirmed by the data and $C^{(2)}_{\rm eff}$ approaches $1$ much
faster than $C^{(1)}_{\rm eff}$, and we present both, for the `S'
data sets of $SU(3)$ with $\beta=14.7172$ and of $SU(6)$ with
$\beta=90.00$, in Fig.~\ref{Cefffig}.

Finally, in our largest statistical sample, which was obtained for $SU(2)$, we find 
\begin{equation}
C^{(2)}_{\rm eff}=0.9991(55) \quad {\rm at} \quad l\simeq 0.68 \,\,{\rm fm}.
\end{equation}

\section{Test of the NG prediction for the 1st excited state}
\label{1st}

\begin{figure}[htb]
\centerline{
\includegraphics[width=8.5cm]{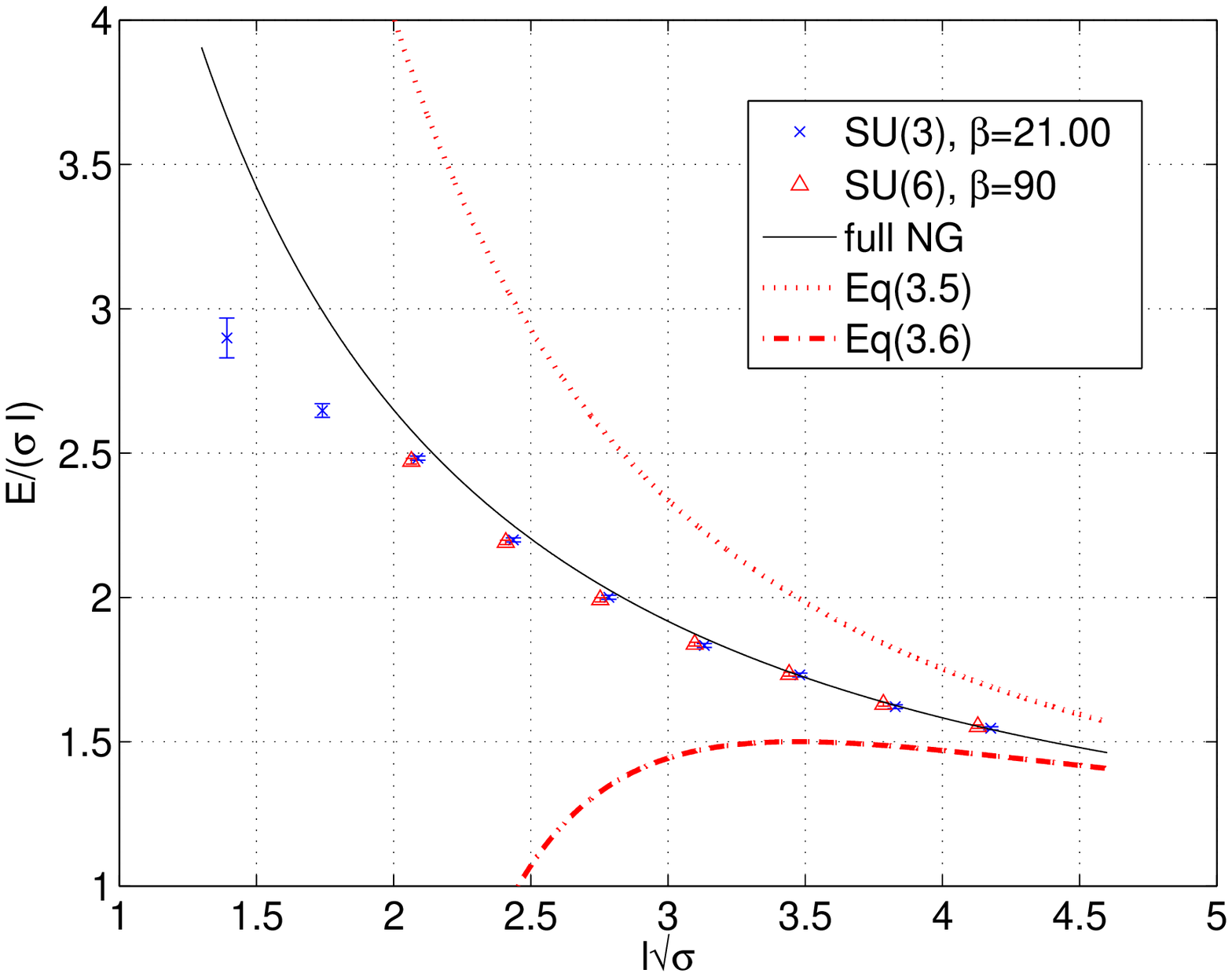} \quad
\includegraphics[width=8.5cm]{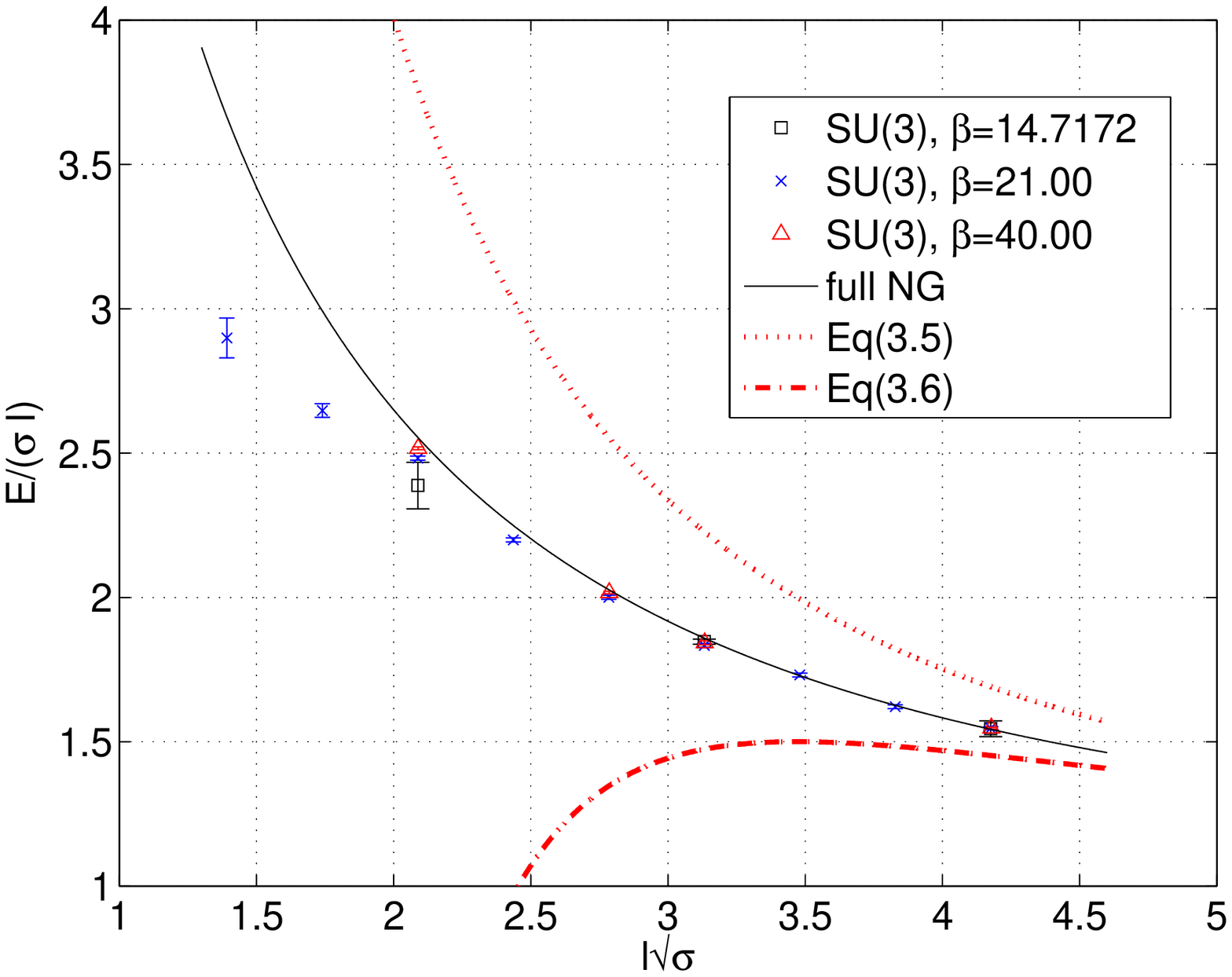} 
}
\caption{The energy of the 1st excited state, divided by $\sigma l$ (See Tables~\ref{fits-gs-partA}-\ref{fits-gs-partB} for the  values of the string tensions). \underline{Left panel:} Comparing $SU(3)$ and $SU(6)$ for a similar lattice spacing of $a^2\sigma \simeq 0.03$.  \underline{Right panel:} Comparing different lattice spacings for $SU(3)$. The black line is the NG prediction, while the dotted(dashed) lines are the NG prediction expanded to leading and next-to-leading order.}
\label{1stfigS}
\end{figure}

Equipped with values for the string tensions, we are now in a
position to test the parameter-free predictions of the NG model and of the Luscher term  formula for the
first excited states. We substitute the values of $a^2\sigma$ that
appear in Table~\ref{fits-gs-partA} in \Eq{NG0} and \Eq{luscher0}
for $w=1, q=0$ and $N_R=N_L=1$ and compare with our 
data. We present the comparison for $SU(3)$ and $SU(6)$, and for two different lattice spacings in Fig.~\ref{1stfigS}. 

 In contrast to the case of the ground state, both the predictions \Eq{luscher0} and \Eq{luscher1} are unable to describe our data for all the values of $l$ that we study, and its seems that they will become consistent with the data only when $l\simeq (6-7)/\sqrt{\sigma}\simeq 2.7-3.2$ fm. Despite this, the NG prediction is strikingly within $\sim 5\%$ of our data for 
 $l\stackrel{>}{_\sim} 2/\sqrt{\sigma} \simeq 0.9$ fm, and becomes consistent when $l\stackrel{>}{_\sim} 3.5/ \sqrt{\sigma}\simeq 1.6$ fm. To further check how this agreement depends on $N$ and whether it evolves with the lattice spacing $a$, we have repeated this analysis for gauge groups with $2\le N\le 8$ and a large set of lattice spacings with $0.2\, {\rm fm} \ge a \ge 0.05$ fm, and string lengths with $l\stackrel{>}{_\sim} 3/\sqrt{\sigma}$, and this will be presented in a forthcoming publication \cite{ABT}. What we find is that the dependence on both $a$ and $N$ is not significant.

\section{Comparison of the NG spectrum with the lowest $\sim 30$ states}
\label{spectrum}

In this section we present results from an extensive calculation
of the lowest $\sim 30$ states in the spectrum. We restrict the discussion
to states with  $w=1$, $q=0,1,2$ and $N_{L,R}\le 3$. We have obtained results for the third excited states with $q=0$ and $N_{L}=N_{R}= 3$ as well, but we postpone their presentation to \cite{ABT}.  The results were obtained for $SU(3)$ with $\beta=21.00,40.00$  and
for $SU(6)$ with $\beta=90.00$. Here we have focused on
strings with $1.4/\sqrt{\sigma} \stackrel{<}{_\sim}
l\stackrel{<}{_\sim} 5.5/\sqrt{\sigma}$.  All energies that we
present here were obtained from single exponential fits.

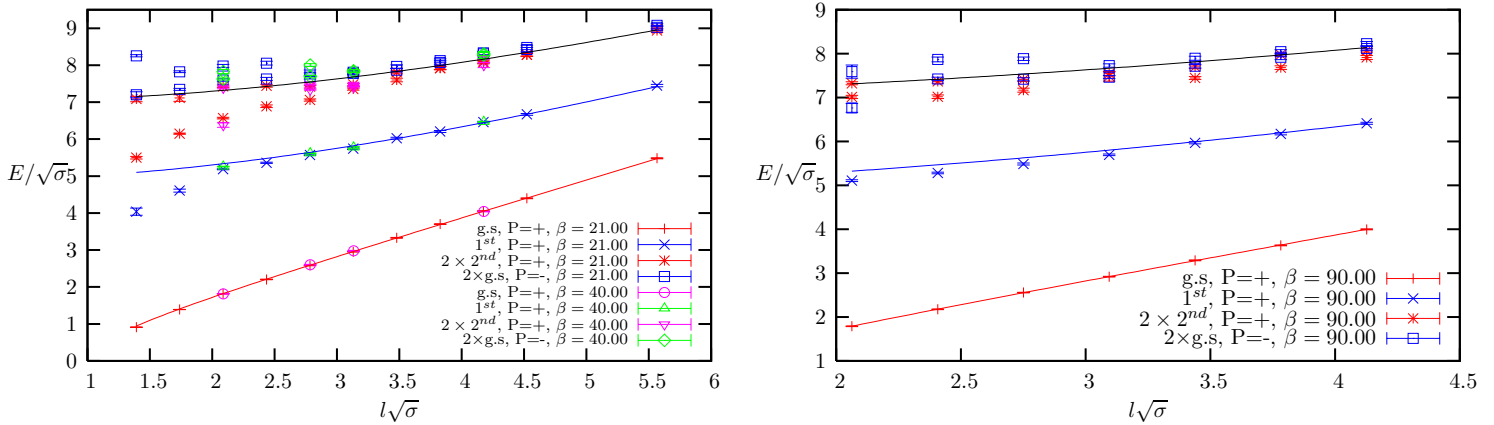
\begin{figure}
\centerline{
\scalebox{0.75}{\input{plotsu3}} 
\, 
\scalebox{0.75}{\input{plotsu62}}
}
\caption{ The energies of the lowest $7$ states in units of $\sqrt{\sigma}$ as a function of $l \sqrt{\sigma}$. The three lines are the NG predictions for the ground state(red), $1^{st}$ excited state(blue) and $2^{nd}$ excited state(black). \underline{Left panel:} We present the results for the case of $SU(3)$ for two different values of $\beta$ (two different lattice spacings). \underline{Right panel:}  Results for $SU(6)$ with $\beta=90.000$. In both panels we denote the degeneracy of the states in the legends.} 
\label{Spectrum}
\end{figure}

\begin{figure}[htb]
\centerline{
\scalebox{1.25}{\input{mom4}}
} 
\caption{$\sqrt{E^2/\sigma - \left( 2 \pi q/ \sqrt{\sigma} l \right)^2}$ as a function of $l\sqrt{\sigma}$ for the ground state (g.s.) and excited states (e.s.) of the non-zero $q$ sector. The four lines present the NG prediction \Eq{NG0}. For $N_R=3, N_L=1$ and $q=2$, the expected degeneracy is three, which agrees with our excited state data, i.e. two(one) states with positive(negative) parity in red(blue).} \label{mom}
\end{figure}
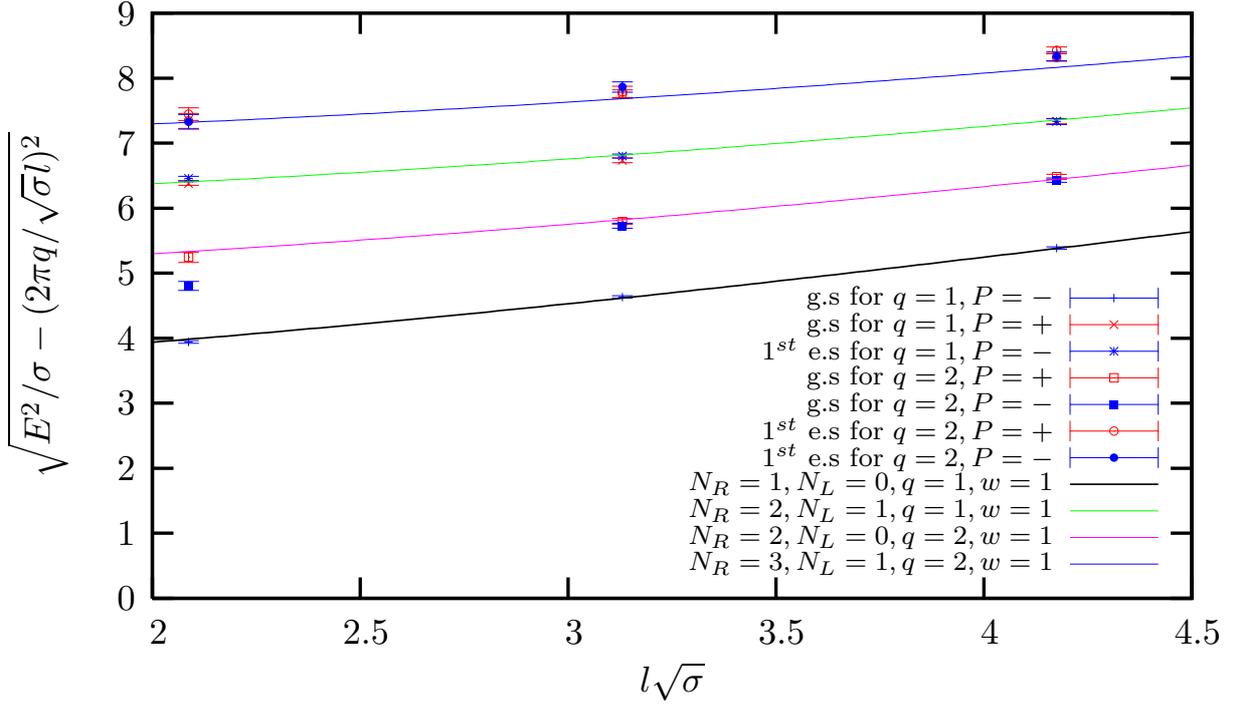

We present the results in Figs.~\ref{Spectrum}-\ref{mom}. The lines are the
predictions of NG model. The string tensions used for these
predictions were extracted from the ground state energies with the
use of the fitting ansatz \Eq{fit}, and we present the fitting
parameters in Table~\ref{fits-gs-partB}.

\begin{table}[htb]
\centering{
\begin{tabular}{|c|c|c|c|c|} \hline \hline
Gauge group  & $\beta$ & $a^2\sigma$ & $C_3$ & Confidence level \\ \hline \hline
\multirow{2}{0.45in}{$SU(3)$} & \,$21.00$ \, & $0.030258(26)$ & \, $0.160(21)$ \, & $69\%$    \\
 & $40.00$  & $0.007577(13)$ & $0.05(31)$ & $15\%$ \\ \hline
$SU(6)$ & $90.00$  & \, $0.029559(36)$ \, & $0.04(21)$   & $88\%$  \\ \hline
\end{tabular}
}
\caption{The parameters $a^2\sigma$ and $C_3$ in the fit \Eq{fit}, which are obtained
  for the $S$ data sets. The errors on $C_3$ are much larger then the ones that appear in Table~\ref{fits-gs-partA} because here we did not study the very short strings.}
\label{fits-gs-partB}
\end{table}

It is clearly seen that the NG predictions are very good approximations to the flux-tube spectrum and deviate from our data only at the level of $\sim 2\%$ once $l\stackrel{>}{_\sim}4.2/\sqrt{\sigma} \simeq 1.9$ fm. At this level some of the systematic errors may be significant. Note that the degeneracy pattern predicted by the NG model is seen from our data : the second energy level is fourfold degenerate at large-$l$. This degeneracy includes two positive parity states and two negative parity states, and these start splitting significantly once $l\stackrel{<}{_\sim}3/\sqrt{\sigma}\simeq 1.35$ fm.

 We note in passing that the $1/l$ and $1/l^3$ contributions to the energies, which were predicted using effective field-theories are clearly insufficient to describe our data, and would presumably do so only for much longer strings. In contrast, the implicit resummation of powers of the Luscher term which appears in the covariant string expression (see discussion at the end of Section~\ref{section_Seff}) is strongly supported by our data.

Next, we fitted the data for the excited states. In the case of the first excited energy level, where there is only one state per level, we used the fitting ansatz  \Eq{fit}. In the case of the second excited energy level, where for each parity there are two states, we used a modified form of \Eq{fit} and fitted the difference between the energies squared of these states to the form
\begin{equation}
\left(\delta E\right)^2 = \sigma \frac{C_p}{\left(l \sqrt{\sigma}\right)^p}.
\end{equation}
 The results of these fits are presented in Table~\ref{fits-p-partB}. 

To compare our data to the Luscher-Weisz prediction in \cite{LW_new} we momentarily assume that $p=2$. Expanding \Eq{fit} this assumption results in an $O(1/l^3)$ contribution to the string energy which deviates from the Luscher-Weisz prediction by an amount proportional to $C_2$. In the case of the first excited state and $l>2/\sqrt{\sigma}$, we substitute the values $C_2\simeq 3-8$ from Table~\ref{fits-p-partB}, and find that this deviation is only at the level of $2\%-6\%$. The smallness of this deviation, and the largeness of the errors we find for $p$ from our fits (see Table~\ref{fits-p-partB}), demonstrate that to unambiguously determine $p$, and test the Luscher-Weisz prediction, requires statistical errors which are at least 2-3 times smaller than the ones our current data has, and a simultaneous control of any systematic errors that may be important at this level of few percents accuracy.  
\begin{table}[htb]
\centering{
\begin{tabular}{|c|c|c|c|c|c|c|} \hline \hline
Level & \, gauge group \, & \, $\beta$ \, & \, $l\sqrt{\sigma}>$ \, & $p$ & $C_p$ & $\chi^2/d.o.f$ \\ \hline \hline
\multirow{4}{0.75in}{$1^{\rm st}$ excited} &  $SU(3)$  & \, $21.00$ \,& $2.1$ & $1.8(5)$ & $6(3)$ & $2.4/7$  \\
&  $SU(3)$  & \, $21.00$ \,& $1.4$ & $3.7(2)$ & $37(5)$ & $39.5/9$  \\
& $SU(3)$   & $40.00$  & $2.1$ & $1.7(7)$ & $3(2)$ & $2.0/2$  \\
& $SU(6)$   & $90.00$  & $2.1$ & $1.6(4)$ & $8(3)$ & $3.8/5$  \\ \hline
\multirow{3}{1.5in}{\quad $2^{\rm nd}$ excited, $P=+$} &  $SU(3)$   & $21.00$  & $2.1$ & $3.0(3)$ & \,$53(15)$ \, & $4.2/7$  \\
& $SU(3)$   & $40.00$  & $2.1$ & $0.5(5)$ & $4(2)$ & $3.2/4$  \\
& $SU(6)$   & $90.00$  & $2.1$ & $0.3(3)$ & $3(1)$ & $3.4/5$  \\ \hline
\multirow{2}{1.5in}{\quad $2^{\rm nd}$ excited, $P=-$} &  $SU(3)$   & $21.00$  & $2.1$ & $2.1(7)$ & $16(9)$ & $7.9/7$  \\
& $SU(6)$   & $90.00$  & $2.1$ & $2.5(5)$ & $36(18)$ & $5.5/5$  \\ \hline
\end{tabular}
}
\caption{The parameter $C_p$ in the fit \Eq{fit} and $\chi^2/d.o.f$, which were obtained for the $S$ data sets (see Section~\ref{contamination}). For the second excited level we fit both the positive parity ($P=+$) and negative parity ($P=-$) states.}
\label{fits-p-partB}
\end{table}

Finally, we move to the nonzero longitudinal momentum $q$ sector. As mentioned in Section~\ref{theory}, in the NG prediction of \Eq{NG0}, the number of left and right movers are constrained to obey the level matching condition \Eq{levels}. The comparison of this prediction with our data is given in Fig.~\ref{mom}, where we present $\sqrt{E^2/\sigma - \left(2\pi q/\sqrt{\sigma}l\right)^2}$ as a function of $l\sqrt{\sigma}$.\footnote{This way of presenting the data 
separates the data sets and eases the representation.} As clearly seen the data is indeed very well described by \Eq{NG0}.

\section{Conclusions}
\label{conclusions}

We have calculated the energy spectrum of closed strings in $SU(N)$ gauge theories in $2+1$ dimensions with the string length $l$ in the range $0.45\,\, {\rm fm}\le l \le 3 \,\,{\rm fm}$. 

For the ground state we have studied $2\le N \le 8$ for lattice spacings $0.06 \le a \le 0.11$ fm, depending on $N$, and saw that the Nambu-Goto (NG) free bosonic string model describes our data very well even at the relatively short distances of $l\simeq 0.6-0.7$ fm (this was already noted in \cite{KN}). In particular, we find that the central charge $c$ is in general consistent with $1$. For example, in $SU(2)$ we see $c=0.9991(55)$ when we extract it already at $l\simeq 0.68$ fm, while in $SU(5)$ we find $c=1.004(23)$ at $l\simeq 0.82$ fm. This provides unambiguous evidence that the closed flux-tube is in the bosonic string universality class.

To study the excited string spectrum we constructed a basis of $\sim 80-200$ operators in each channel, and from a variational calculation we extracted the lowest $\sim 30$ states. These include states with positive and negative parity, as well as with nonzero momentum along the string direction. In general we find that the agreement with the NG prediction in \Eq{NG0} is very good, including the expected degeneracy pattern. 

This agreement is in striking contrast to what we find when we simply compare to the Luscher term as in \Eq{luscher0} or to the Luscher-Weisz prediction in \Eq{luscher1}. While for the ground states these describe our data already at $l\simeq 1.35$ fm, where they are indistinguishable from the full NG formula, for the excited states the situation is completely different. In particular, whereas the NG prediction works well for the first excited state already at $l\simeq 1.35$ fm, the predictions of Eqs.~(\ref{luscher0}) and (\ref{luscher1}) are still very far from the data. 

This means that our results show unambiguously that the confining flux-tube can be described by a covariant string theory with small/moderate corrections, down to very short distance scales, and possibly at all distance scales at large-$N$.

\section*{Acknowledgements}

AA acknowledges the support of the EC 6$^{th}$ Framework Programme Research
and Training Network MRTN-CT-2004-503369. BB was supported by PPARC and thanks the Isaac Newton Institute for Mathematical Sciences at Cambridge, in which this work was completed. The computations were performed on machines
funded primarily by Oxford and EPSRC.

\end{document}

%% file: plotsu3.tex
\begingroup%
  \makeatletter%
  \newcommand{\GNUPLOTspecial}{%
    \@sanitize\catcode`\%=14\relax\special}%
  \setlength{\unitlength}{0.1bp}%
\begin{picture}(3600,2160)(0,0)%
{\GNUPLOTspecial{"
/gnudict 256 dict def
gnudict begin
/Color true def
/Solid true def
/gnulinewidth 5.000 def
/userlinewidth gnulinewidth def
/vshift -33 def
/dl {10.0 mul} def
/hpt_ 31.5 def
/vpt_ 31.5 def
/hpt hpt_ def
/vpt vpt_ def
/Rounded false def
/M {moveto} bind def
/L {lineto} bind def
/R {rmoveto} bind def
/V {rlineto} bind def
/N {newpath moveto} bind def
/C {setrgbcolor} bind def
/f {rlineto fill} bind def
/vpt2 vpt 2 mul def
/hpt2 hpt 2 mul def
/Lshow { currentpoint stroke M
  0 vshift R show } def
/Rshow { currentpoint stroke M
  dup stringwidth pop neg vshift R show } def
/Cshow { currentpoint stroke M
  dup stringwidth pop -2 div vshift R show } def
/UP { dup vpt_ mul /vpt exch def hpt_ mul /hpt exch def
  /hpt2 hpt 2 mul def /vpt2 vpt 2 mul def } def
/DL { Color {setrgbcolor Solid {pop []} if 0 setdash }
 {pop pop pop 0 setgray Solid {pop []} if 0 setdash} ifelse } def
/BL { stroke userlinewidth 2 mul setlinewidth
      Rounded { 1 setlinejoin 1 setlinecap } if } def
/AL { stroke userlinewidth 2 div setlinewidth
      Rounded { 1 setlinejoin 1 setlinecap } if } def
/UL { dup gnulinewidth mul /userlinewidth exch def
      dup 1 lt {pop 1} if 10 mul /udl exch def } def
/PL { stroke userlinewidth setlinewidth
      Rounded { 1 setlinejoin 1 setlinecap } if } def
/LTw { PL [] 1 setgray } def
/LTb { BL [] 0 0 0 DL } def
/LTa { AL [1 udl mul 2 udl mul] 0 setdash 0 0 0 setrgbcolor } def
/LT0 { PL [] 1 0 0 DL } def
/LT1 { PL [4 dl 2 dl] 0 1 0 DL } def
/LT2 { PL [2 dl 3 dl] 0 0 1 DL } def
/LT3 { PL [1 dl 1.5 dl] 1 0 1 DL } def
/LT4 { PL [5 dl 2 dl 1 dl 2 dl] 0 1 1 DL } def
/LT5 { PL [4 dl 3 dl 1 dl 3 dl] 1 1 0 DL } def
/LT6 { PL [2 dl 2 dl 2 dl 4 dl] 0 0 0 DL } def
/LT7 { PL [2 dl 2 dl 2 dl 2 dl 2 dl 4 dl] 1 0.3 0 DL } def
/LT8 { PL [2 dl 2 dl 2 dl 2 dl 2 dl 2 dl 2 dl 4 dl] 0.5 0.5 0.5 DL } def
/Pnt { stroke [] 0 setdash
   gsave 1 setlinecap M 0 0 V stroke grestore } def
/Dia { stroke [] 0 setdash 2 copy vpt add M
  hpt neg vpt neg V hpt vpt neg V
  hpt vpt V hpt neg vpt V closepath stroke
  Pnt } def
/Pls { stroke [] 0 setdash vpt sub M 0 vpt2 V
  currentpoint stroke M
  hpt neg vpt neg R hpt2 0 V stroke
  } def
/Box { stroke [] 0 setdash 2 copy exch hpt sub exch vpt add M
  0 vpt2 neg V hpt2 0 V 0 vpt2 V
  hpt2 neg 0 V closepath stroke
  Pnt } def
/Crs { stroke [] 0 setdash exch hpt sub exch vpt add M
  hpt2 vpt2 neg V currentpoint stroke M
  hpt2 neg 0 R hpt2 vpt2 V stroke } def
/TriU { stroke [] 0 setdash 2 copy vpt 1.12 mul add M
  hpt neg vpt -1.62 mul V
  hpt 2 mul 0 V
  hpt neg vpt 1.62 mul V closepath stroke
  Pnt  } def
/Star { 2 copy Pls Crs } def
/BoxF { stroke [] 0 setdash exch hpt sub exch vpt add M
  0 vpt2 neg V  hpt2 0 V  0 vpt2 V
  hpt2 neg 0 V  closepath fill } def
/TriUF { stroke [] 0 setdash vpt 1.12 mul add M
  hpt neg vpt -1.62 mul V
  hpt 2 mul 0 V
  hpt neg vpt 1.62 mul V closepath fill } def
/TriD { stroke [] 0 setdash 2 copy vpt 1.12 mul sub M
  hpt neg vpt 1.62 mul V
  hpt 2 mul 0 V
  hpt neg vpt -1.62 mul V closepath stroke
  Pnt  } def
/TriDF { stroke [] 0 setdash vpt 1.12 mul sub M
  hpt neg vpt 1.62 mul V
  hpt 2 mul 0 V
  hpt neg vpt -1.62 mul V closepath fill} def
/DiaF { stroke [] 0 setdash vpt add M
  hpt neg vpt neg V hpt vpt neg V
  hpt vpt V hpt neg vpt V closepath fill } def
/Pent { stroke [] 0 setdash 2 copy gsave
  translate 0 hpt M 4 {72 rotate 0 hpt L} repeat
  closepath stroke grestore Pnt } def
/PentF { stroke [] 0 setdash gsave
  translate 0 hpt M 4 {72 rotate 0 hpt L} repeat
  closepath fill grestore } def
/Circle { stroke [] 0 setdash 2 copy
  hpt 0 360 arc stroke Pnt } def
/CircleF { stroke [] 0 setdash hpt 0 360 arc fill } def
/C0 { BL [] 0 setdash 2 copy moveto vpt 90 450  arc } bind def
/C1 { BL [] 0 setdash 2 copy        moveto
       2 copy  vpt 0 90 arc closepath fill
               vpt 0 360 arc closepath } bind def
/C2 { BL [] 0 setdash 2 copy moveto
       2 copy  vpt 90 180 arc closepath fill
               vpt 0 360 arc closepath } bind def
/C3 { BL [] 0 setdash 2 copy moveto
       2 copy  vpt 0 180 arc closepath fill
               vpt 0 360 arc closepath } bind def
/C4 { BL [] 0 setdash 2 copy moveto
       2 copy  vpt 180 270 arc closepath fill
               vpt 0 360 arc closepath } bind def
/C5 { BL [] 0 setdash 2 copy moveto
       2 copy  vpt 0 90 arc
       2 copy moveto
       2 copy  vpt 180 270 arc closepath fill
               vpt 0 360 arc } bind def
/C6 { BL [] 0 setdash 2 copy moveto
      2 copy  vpt 90 270 arc closepath fill
              vpt 0 360 arc closepath } bind def
/C7 { BL [] 0 setdash 2 copy moveto
      2 copy  vpt 0 270 arc closepath fill
              vpt 0 360 arc closepath } bind def
/C8 { BL [] 0 setdash 2 copy moveto
      2 copy vpt 270 360 arc closepath fill
              vpt 0 360 arc closepath } bind def
/C9 { BL [] 0 setdash 2 copy moveto
      2 copy  vpt 270 450 arc closepath fill
              vpt 0 360 arc closepath } bind def
/C10 { BL [] 0 setdash 2 copy 2 copy moveto vpt 270 360 arc closepath fill
       2 copy moveto
       2 copy vpt 90 180 arc closepath fill
               vpt 0 360 arc closepath } bind def
/C11 { BL [] 0 setdash 2 copy moveto
       2 copy  vpt 0 180 arc closepath fill
       2 copy moveto
       2 copy  vpt 270 360 arc closepath fill
               vpt 0 360 arc closepath } bind def
/C12 { BL [] 0 setdash 2 copy moveto
       2 copy  vpt 180 360 arc closepath fill
               vpt 0 360 arc closepath } bind def
/C13 { BL [] 0 setdash  2 copy moveto
       2 copy  vpt 0 90 arc closepath fill
       2 copy moveto
       2 copy  vpt 180 360 arc closepath fill
               vpt 0 360 arc closepath } bind def
/C14 { BL [] 0 setdash 2 copy moveto
       2 copy  vpt 90 360 arc closepath fill
               vpt 0 360 arc } bind def
/C15 { BL [] 0 setdash 2 copy vpt 0 360 arc closepath fill
               vpt 0 360 arc closepath } bind def
/Rec   { newpath 4 2 roll moveto 1 index 0 rlineto 0 exch rlineto
       neg 0 rlineto closepath } bind def
/Square { dup Rec } bind def
/Bsquare { vpt sub exch vpt sub exch vpt2 Square } bind def
/S0 { BL [] 0 setdash 2 copy moveto 0 vpt rlineto BL Bsquare } bind def
/S1 { BL [] 0 setdash 2 copy vpt Square fill Bsquare } bind def
/S2 { BL [] 0 setdash 2 copy exch vpt sub exch vpt Square fill Bsquare } bind def
/S3 { BL [] 0 setdash 2 copy exch vpt sub exch vpt2 vpt Rec fill Bsquare } bind def
/S4 { BL [] 0 setdash 2 copy exch vpt sub exch vpt sub vpt Square fill Bsquare } bind def
/S5 { BL [] 0 setdash 2 copy 2 copy vpt Square fill
       exch vpt sub exch vpt sub vpt Square fill Bsquare } bind def
/S6 { BL [] 0 setdash 2 copy exch vpt sub exch vpt sub vpt vpt2 Rec fill Bsquare } bind def
/S7 { BL [] 0 setdash 2 copy exch vpt sub exch vpt sub vpt vpt2 Rec fill
       2 copy vpt Square fill
       Bsquare } bind def
/S8 { BL [] 0 setdash 2 copy vpt sub vpt Square fill Bsquare } bind def
/S9 { BL [] 0 setdash 2 copy vpt sub vpt vpt2 Rec fill Bsquare } bind def
/S10 { BL [] 0 setdash 2 copy vpt sub vpt Square fill 2 copy exch vpt sub exch vpt Square fill
       Bsquare } bind def
/S11 { BL [] 0 setdash 2 copy vpt sub vpt Square fill 2 copy exch vpt sub exch vpt2 vpt Rec fill
       Bsquare } bind def
/S12 { BL [] 0 setdash 2 copy exch vpt sub exch vpt sub vpt2 vpt Rec fill Bsquare } bind def
/S13 { BL [] 0 setdash 2 copy exch vpt sub exch vpt sub vpt2 vpt Rec fill
       2 copy vpt Square fill Bsquare } bind def
/S14 { BL [] 0 setdash 2 copy exch vpt sub exch vpt sub vpt2 vpt Rec fill
       2 copy exch vpt sub exch vpt Square fill Bsquare } bind def
/S15 { BL [] 0 setdash 2 copy Bsquare fill Bsquare } bind def
/D0 { gsave translate 45 rotate 0 0 S0 stroke grestore } bind def
/D1 { gsave translate 45 rotate 0 0 S1 stroke grestore } bind def
/D2 { gsave translate 45 rotate 0 0 S2 stroke grestore } bind def
/D3 { gsave translate 45 rotate 0 0 S3 stroke grestore } bind def
/D4 { gsave translate 45 rotate 0 0 S4 stroke grestore } bind def
/D5 { gsave translate 45 rotate 0 0 S5 stroke grestore } bind def
/D6 { gsave translate 45 rotate 0 0 S6 stroke grestore } bind def
/D7 { gsave translate 45 rotate 0 0 S7 stroke grestore } bind def
/D8 { gsave translate 45 rotate 0 0 S8 stroke grestore } bind def
/D9 { gsave translate 45 rotate 0 0 S9 stroke grestore } bind def
/D10 { gsave translate 45 rotate 0 0 S10 stroke grestore } bind def
/D11 { gsave translate 45 rotate 0 0 S11 stroke grestore } bind def
/D12 { gsave translate 45 rotate 0 0 S12 stroke grestore } bind def
/D13 { gsave translate 45 rotate 0 0 S13 stroke grestore } bind def
/D14 { gsave translate 45 rotate 0 0 S14 stroke grestore } bind def
/D15 { gsave translate 45 rotate 0 0 S15 stroke grestore } bind def
/DiaE { stroke [] 0 setdash vpt add M
  hpt neg vpt neg V hpt vpt neg V
  hpt vpt V hpt neg vpt V closepath stroke } def
/BoxE { stroke [] 0 setdash exch hpt sub exch vpt add M
  0 vpt2 neg V hpt2 0 V 0 vpt2 V
  hpt2 neg 0 V closepath stroke } def
/TriUE { stroke [] 0 setdash vpt 1.12 mul add M
  hpt neg vpt -1.62 mul V
  hpt 2 mul 0 V
  hpt neg vpt 1.62 mul V closepath stroke } def
/TriDE { stroke [] 0 setdash vpt 1.12 mul sub M
  hpt neg vpt 1.62 mul V
  hpt 2 mul 0 V
  hpt neg vpt -1.62 mul V closepath stroke } def
/PentE { stroke [] 0 setdash gsave
  translate 0 hpt M 4 {72 rotate 0 hpt L} repeat
  closepath stroke grestore } def
/CircE { stroke [] 0 setdash 
  hpt 0 360 arc stroke } def
/Opaque { gsave closepath 1 setgray fill grestore 0 setgray closepath } def
/DiaW { stroke [] 0 setdash vpt add M
  hpt neg vpt neg V hpt vpt neg V
  hpt vpt V hpt neg vpt V Opaque stroke } def
/BoxW { stroke [] 0 setdash exch hpt sub exch vpt add M
  0 vpt2 neg V hpt2 0 V 0 vpt2 V
  hpt2 neg 0 V Opaque stroke } def
/TriUW { stroke [] 0 setdash vpt 1.12 mul add M
  hpt neg vpt -1.62 mul V
  hpt 2 mul 0 V
  hpt neg vpt 1.62 mul V Opaque stroke } def
/TriDW { stroke [] 0 setdash vpt 1.12 mul sub M
  hpt neg vpt 1.62 mul V
  hpt 2 mul 0 V
  hpt neg vpt -1.62 mul V Opaque stroke } def
/PentW { stroke [] 0 setdash gsave
  translate 0 hpt M 4 {72 rotate 0 hpt L} repeat
  Opaque stroke grestore } def
/CircW { stroke [] 0 setdash 
  hpt 0 360 arc Opaque stroke } def
/BoxFill { gsave Rec 1 setgray fill grestore } def
/BoxColFill {
  gsave Rec
  /Fillden exch def
  currentrgbcolor
  /ColB exch def /ColG exch def /ColR exch def
  /ColR ColR Fillden mul Fillden sub 1 add def
  /ColG ColG Fillden mul Fillden sub 1 add def
  /ColB ColB Fillden mul Fillden sub 1 add def
  ColR ColG ColB setrgbcolor
  fill grestore } def
%
%
/PatternFill { gsave /PFa [ 9 2 roll ] def
    PFa 0 get PFa 2 get 2 div add PFa 1 get PFa 3 get 2 div add translate
    PFa 2 get -2 div PFa 3 get -2 div PFa 2 get PFa 3 get Rec
    gsave 1 setgray fill grestore clip
    currentlinewidth 0.5 mul setlinewidth
    /PFs PFa 2 get dup mul PFa 3 get dup mul add sqrt def
    0 0 M PFa 5 get rotate PFs -2 div dup translate
	0 1 PFs PFa 4 get div 1 add floor cvi
	{ PFa 4 get mul 0 M 0 PFs V } for
    0 PFa 6 get ne {
	0 1 PFs PFa 4 get div 1 add floor cvi
	{ PFa 4 get mul 0 2 1 roll M PFs 0 V } for
    } if
    stroke grestore } def
/Symbol-Oblique /Symbol findfont [1 0 .167 1 0 0] makefont
dup length dict begin {1 index /FID eq {pop pop} {def} ifelse} forall
currentdict end definefont pop
end
gnudict begin
gsave
0 0 translate
0.100 0.100 scale
0 setgray
newpath
1.000 UL
LTb
350 300 M
63 0 V
3037 0 R
-63 0 V
1.000 UL
LTb
350 485 M
63 0 V
3037 0 R
-63 0 V
1.000 UL
LTb
350 671 M
63 0 V
3037 0 R
-63 0 V
1.000 UL
LTb
350 856 M
63 0 V
3037 0 R
-63 0 V
1.000 UL
LTb
350 1041 M
63 0 V
3037 0 R
-63 0 V
1.000 UL
LTb
350 1226 M
63 0 V
3037 0 R
-63 0 V
1.000 UL
LTb
350 1412 M
63 0 V
3037 0 R
-63 0 V
1.000 UL
LTb
350 1597 M
63 0 V
3037 0 R
-63 0 V
1.000 UL
LTb
350 1782 M
63 0 V
3037 0 R
-63 0 V
1.000 UL
LTb
350 1967 M
63 0 V
3037 0 R
-63 0 V
1.000 UL
LTb
350 300 M
0 63 V
0 1697 R
0 -63 V
1.000 UL
LTb
660 300 M
0 63 V
0 1697 R
0 -63 V
1.000 UL
LTb
970 300 M
0 63 V
0 1697 R
0 -63 V
1.000 UL
LTb
1280 300 M
0 63 V
0 1697 R
0 -63 V
1.000 UL
LTb
1590 300 M
0 63 V
0 1697 R
0 -63 V
1.000 UL
LTb
1900 300 M
0 63 V
0 1697 R
0 -63 V
1.000 UL
LTb
2210 300 M
0 63 V
0 1697 R
0 -63 V
1.000 UL
LTb
2520 300 M
0 63 V
0 1697 R
0 -63 V
1.000 UL
LTb
2830 300 M
0 63 V
0 1697 R
0 -63 V
1.000 UL
LTb
3140 300 M
0 63 V
0 1697 R
0 -63 V
1.000 UL
LTb
3450 300 M
0 63 V
0 1697 R
0 -63 V
1.000 UL
LTb
1.000 UL
LTb
350 300 M
3100 0 V
0 1760 V
-3100 0 V
350 300 L
LTb
LTb
1.000 UP
0.800 UP
0.500 UL
LT0
LTb
LT0
3087 963 M
263 0 V
-263 31 R
0 -62 V
263 62 R
0 -62 V
593 468 M
0 2 V
-31 -2 R
62 0 V
-62 2 R
62 0 V
184 87 R
0 1 V
-31 -1 R
62 0 V
-62 1 R
62 0 V
185 78 R
0 2 V
-31 -2 R
62 0 V
-62 2 R
62 0 V
185 69 R
0 3 V
-31 -3 R
62 0 V
-62 3 R
62 0 V
185 67 R
0 3 V
-31 -3 R
62 0 V
-62 3 R
62 0 V
184 66 R
0 3 V
-31 -3 R
62 0 V
-62 3 R
62 0 V
185 66 R
0 4 V
-31 -4 R
62 0 V
-62 4 R
62 0 V
185 65 R
0 3 V
-31 -3 R
62 0 V
-62 3 R
62 0 V
184 62 R
0 4 V
-31 -4 R
62 0 V
-62 4 R
62 0 V
185 60 R
0 4 V
-31 -4 R
62 0 V
-62 4 R
62 0 V
616 196 R
0 5 V
-31 -5 R
62 0 V
-62 5 R
62 0 V
593 469 Pls
808 558 Pls
1024 637 Pls
1240 708 Pls
1456 779 Pls
1671 847 Pls
1887 917 Pls
2103 986 Pls
2318 1051 Pls
2534 1115 Pls
3181 1316 Pls
3218 963 Pls
0.800 UP
0.500 UL
LT2
LTb
LT2
3087 883 M
263 0 V
-263 31 R
0 -62 V
263 62 R
0 -62 V
593 1030 M
0 36 V
-31 -36 R
62 0 V
-62 36 R
62 0 V
184 80 R
0 15 V
-31 -15 R
62 0 V
-62 15 R
62 0 V
185 97 R
0 5 V
-31 -5 R
62 0 V
-62 5 R
62 0 V
185 27 R
0 6 V
-31 -6 R
62 0 V
-62 6 R
62 0 V
185 33 R
0 7 V
-31 -7 R
62 0 V
-62 7 R
62 0 V
184 24 R
0 8 V
-31 -8 R
62 0 V
-62 8 R
62 0 V
185 44 R
0 9 V
-31 -9 R
62 0 V
-62 9 R
62 0 V
185 24 R
0 10 V
-31 -10 R
62 0 V
-62 10 R
62 0 V
184 38 R
0 8 V
-31 -8 R
62 0 V
-62 8 R
62 0 V
185 30 R
0 9 V
-31 -9 R
62 0 V
-62 9 R
62 0 V
616 134 R
0 12 V
-31 -12 R
62 0 V
-62 12 R
62 0 V
593 1048 Crs
808 1154 Crs
1024 1261 Crs
1240 1293 Crs
1456 1332 Crs
1671 1364 Crs
1887 1416 Crs
2103 1450 Crs
2318 1497 Crs
2534 1536 Crs
3181 1680 Crs
3218 883 Crs
0.800 UP
0.500 UL
LT0
LTb
LT0
3087 803 M
263 0 V
-263 31 R
0 -62 V
263 62 R
0 -62 V
593 1312 M
0 13 V
-31 -13 R
62 0 V
-62 13 R
62 0 V
-31 274 R
0 27 V
-31 -27 R
62 0 V
-62 27 R
62 0 V
808 1436 M
0 6 V
-31 -6 R
62 0 V
-62 6 R
62 0 V
-31 159 R
0 33 V
-31 -33 R
62 0 V
-62 33 R
62 0 V
185 -121 R
0 8 V
-31 -8 R
62 0 V
-62 8 R
62 0 V
-31 149 R
0 13 V
-31 -13 R
62 0 V
-62 13 R
62 0 V
185 -113 R
0 13 V
-31 -13 R
62 0 V
-62 13 R
62 0 V
-31 89 R
0 14 V
-31 -14 R
62 0 V
-62 14 R
62 0 V
185 -84 R
0 11 V
-31 -11 R
62 0 V
-62 11 R
62 0 V
-31 55 R
0 13 V
-31 -13 R
62 0 V
-62 13 R
62 0 V
184 -24 R
0 11 V
-31 -11 R
62 0 V
-62 11 R
62 0 V
-31 11 R
0 13 V
-31 -13 R
62 0 V
-62 13 R
62 0 V
185 7 R
0 18 V
-31 -18 R
62 0 V
-62 18 R
62 0 V
-31 22 R
0 16 V
-31 -16 R
62 0 V
-62 16 R
62 0 V
185 4 R
0 16 V
-31 -16 R
62 0 V
-62 16 R
62 0 V
-31 -14 R
0 15 V
-31 -15 R
62 0 V
-62 15 R
62 0 V
184 13 R
0 14 V
-31 -14 R
62 0 V
stroke
2287 1803 M
62 0 V
-31 -11 R
0 17 V
-31 -17 R
62 0 V
-62 17 R
62 0 V
185 18 R
0 15 V
-31 -15 R
62 0 V
-62 15 R
62 0 V
-31 -15 R
0 15 V
-31 -15 R
62 0 V
-62 15 R
62 0 V
616 104 R
0 19 V
-31 -19 R
62 0 V
-62 19 R
62 0 V
-31 -8 R
0 19 V
-31 -19 R
62 0 V
-62 19 R
62 0 V
593 1319 Star
593 1613 Star
808 1439 Star
808 1618 Star
1024 1517 Star
1024 1676 Star
1240 1576 Star
1240 1679 Star
1456 1608 Star
1456 1675 Star
1671 1663 Star
1671 1686 Star
1887 1708 Star
1887 1747 Star
2103 1767 Star
2103 1768 Star
2318 1796 Star
2318 1801 Star
2534 1834 Star
2534 1835 Star
3181 1956 Star
3181 1966 Star
3218 803 Star
0.800 UP
0.500 UL
LT2
LTb
LT2
3087 723 M
263 0 V
-263 31 R
0 -62 V
263 62 R
0 -62 V
593 1619 M
0 30 V
-31 -30 R
62 0 V
-62 30 R
62 0 V
-31 175 R
0 11 V
-31 -11 R
62 0 V
-62 11 R
62 0 V
808 1656 M
0 10 V
-31 -10 R
62 0 V
-62 10 R
62 0 V
-31 79 R
0 9 V
-31 -9 R
62 0 V
-62 9 R
62 0 V
185 -52 R
0 12 V
-31 -12 R
62 0 V
-62 12 R
62 0 V
-31 57 R
0 15 V
-31 -15 R
62 0 V
-62 15 R
62 0 V
185 -80 R
0 15 V
-31 -15 R
62 0 V
-62 15 R
62 0 V
-31 62 R
0 19 V
-31 -19 R
62 0 V
-62 19 R
62 0 V
185 -89 R
0 19 V
-31 -19 R
62 0 V
-62 19 R
62 0 V
-31 12 R
0 14 V
-31 -14 R
62 0 V
-62 14 R
62 0 V
184 -20 R
0 18 V
-31 -18 R
62 0 V
-62 18 R
62 0 V
-31 -23 R
0 16 V
-31 -16 R
62 0 V
-62 16 R
62 0 V
185 0 R
0 16 V
-31 -16 R
62 0 V
-62 16 R
62 0 V
-31 2 R
0 17 V
-31 -17 R
62 0 V
-62 17 R
62 0 V
185 3 R
0 18 V
-31 -18 R
62 0 V
-62 18 R
62 0 V
-31 -8 R
0 17 V
-31 -17 R
62 0 V
-62 17 R
62 0 V
184 16 R
0 18 V
-31 -18 R
62 0 V
stroke
2287 1848 M
62 0 V
-31 -13 R
0 17 V
-31 -17 R
62 0 V
-62 17 R
62 0 V
185 -2 R
0 21 V
-31 -21 R
62 0 V
-62 21 R
62 0 V
-31 -11 R
0 22 V
-31 -22 R
62 0 V
-62 22 R
62 0 V
616 82 R
0 17 V
-31 -17 R
62 0 V
-62 17 R
62 0 V
-31 -6 R
0 15 V
-31 -15 R
62 0 V
-62 15 R
62 0 V
593 1634 Box
593 1829 Box
808 1661 Box
808 1749 Box
1024 1708 Box
1024 1778 Box
1240 1713 Box
1240 1792 Box
1456 1722 Box
1456 1751 Box
1671 1747 Box
1671 1741 Box
1887 1757 Box
1887 1776 Box
2103 1796 Box
2103 1805 Box
2318 1839 Box
2318 1844 Box
2534 1861 Box
2534 1871 Box
3181 1972 Box
3181 1982 Box
3218 723 Box
0.800 UP
0.500 UL
LT3
LTb
LT3
3087 643 M
263 0 V
-263 31 R
0 -62 V
263 62 R
0 -62 V
1025 635 M
0 3 V
-31 -3 R
62 0 V
-62 3 R
62 0 V
401 143 R
0 3 V
-31 -3 R
62 0 V
-62 3 R
62 0 V
185 65 R
0 5 V
-31 -5 R
62 0 V
-62 5 R
62 0 V
616 193 R
0 3 V
-31 -3 R
62 0 V
-62 3 R
62 0 V
1025 636 Circle
1457 782 Circle
1673 852 Circle
2320 1049 Circle
3218 643 Circle
0.800 UP
0.500 UL
LT1
LTb
LT1
3087 563 M
263 0 V
-263 31 R
0 -62 V
263 62 R
0 -62 V
1025 1272 M
0 4 V
-31 -4 R
62 0 V
-62 4 R
62 0 V
401 62 R
0 5 V
-31 -5 R
62 0 V
-62 5 R
62 0 V
185 23 R
0 8 V
-31 -8 R
62 0 V
-62 8 R
62 0 V
616 122 R
0 4 V
-31 -4 R
62 0 V
-62 4 R
62 0 V
1025 1274 TriU
1457 1341 TriU
1673 1370 TriU
2320 1498 TriU
3218 563 TriU
0.800 UP
0.500 UL
LT3
LTb
LT3
3087 483 M
263 0 V
-263 31 R
0 -62 V
263 62 R
0 -62 V
1025 1471 M
0 22 V
-31 -22 R
62 0 V
-62 22 R
62 0 V
-31 171 R
0 10 V
-31 -10 R
62 0 V
-62 10 R
62 0 V
401 -17 R
0 6 V
-31 -6 R
62 0 V
-62 6 R
62 0 V
-31 19 R
0 6 V
-31 -6 R
62 0 V
-62 6 R
62 0 V
185 -20 R
0 8 V
-31 -8 R
62 0 V
-62 8 R
62 0 V
-31 2 R
0 12 V
-31 -12 R
62 0 V
-62 12 R
62 0 V
616 123 R
0 12 V
-31 -12 R
62 0 V
-62 12 R
62 0 V
-31 -55 R
0 26 V
-31 -26 R
62 0 V
-62 26 R
62 0 V
1025 1482 TriD
1025 1669 TriD
1457 1660 TriD
1457 1685 TriD
1673 1672 TriD
1673 1684 TriD
2320 1819 TriD
2320 1783 TriD
3218 483 TriD
1.000 UP
0.500 UL
LT1
LTb
LT1
3087 403 M
263 0 V
-263 31 R
0 -62 V
263 62 R
0 -62 V
1025 1708 M
0 10 V
-31 -10 R
62 0 V
-62 10 R
62 0 V
-31 22 R
0 21 V
-31 -21 R
62 0 V
-62 21 R
62 0 V
401 -42 R
0 12 V
-31 -12 R
62 0 V
-62 12 R
62 0 V
-31 45 R
0 14 V
-31 -14 R
62 0 V
-62 14 R
62 0 V
185 -45 R
0 8 V
-31 -8 R
62 0 V
-62 8 R
62 0 V
-31 -7 R
0 15 V
-31 -15 R
62 0 V
-62 15 R
62 0 V
616 70 R
0 16 V
-31 -16 R
62 0 V
-62 16 R
62 0 V
-31 -31 R
0 34 V
-31 -34 R
62 0 V
-62 34 R
62 0 V
1025 1713 Dia
1025 1750 Dia
1457 1725 Dia
1457 1783 Dia
1673 1749 Dia
1673 1754 Dia
2320 1839 Dia
2320 1833 Dia
3218 403 Dia
0.200 UL
LT0
593 475 M
26 11 V
26 11 V
26 11 V
26 10 V
27 10 V
26 11 V
26 9 V
26 10 V
26 10 V
26 9 V
26 10 V
27 9 V
26 9 V
26 9 V
26 10 V
26 9 V
26 9 V
26 8 V
27 9 V
26 9 V
26 9 V
26 9 V
26 8 V
26 9 V
26 9 V
27 8 V
26 9 V
26 8 V
26 9 V
26 8 V
26 9 V
26 8 V
27 8 V
26 9 V
26 8 V
26 9 V
26 8 V
26 8 V
26 9 V
27 8 V
26 8 V
26 8 V
26 9 V
26 8 V
26 8 V
26 8 V
27 9 V
26 8 V
26 8 V
26 8 V
26 8 V
26 8 V
26 9 V
27 8 V
26 8 V
26 8 V
26 8 V
26 8 V
26 8 V
26 8 V
27 8 V
26 9 V
26 8 V
26 8 V
26 8 V
26 8 V
26 8 V
27 8 V
26 8 V
26 8 V
26 8 V
26 8 V
26 8 V
26 8 V
27 8 V
26 8 V
26 8 V
26 8 V
26 8 V
26 8 V
27 8 V
26 8 V
26 8 V
26 8 V
26 8 V
26 8 V
26 8 V
27 8 V
26 8 V
26 8 V
26 8 V
26 8 V
26 8 V
26 8 V
27 8 V
26 8 V
26 8 V
26 8 V
26 8 V
0.200 UL
LT2
593 1245 M
26 2 V
26 2 V
26 3 V
26 2 V
27 2 V
26 3 V
26 2 V
26 3 V
26 3 V
26 2 V
26 3 V
27 3 V
26 3 V
26 3 V
26 3 V
26 3 V
26 3 V
26 3 V
27 3 V
26 3 V
26 3 V
26 4 V
26 3 V
26 3 V
26 4 V
27 3 V
26 4 V
26 4 V
26 3 V
26 4 V
26 4 V
26 3 V
27 4 V
26 4 V
26 4 V
26 4 V
26 4 V
26 4 V
26 4 V
27 4 V
26 4 V
26 5 V
26 4 V
26 4 V
26 5 V
26 4 V
27 4 V
26 5 V
26 4 V
26 5 V
26 4 V
26 5 V
26 5 V
27 4 V
26 5 V
26 5 V
26 4 V
26 5 V
26 5 V
26 5 V
27 5 V
26 5 V
26 5 V
26 5 V
26 5 V
26 5 V
26 5 V
27 5 V
26 5 V
26 5 V
26 5 V
26 6 V
26 5 V
26 5 V
27 5 V
26 6 V
26 5 V
26 5 V
26 6 V
26 5 V
27 6 V
26 5 V
26 6 V
26 5 V
26 6 V
26 5 V
26 6 V
27 6 V
26 5 V
26 6 V
26 6 V
26 5 V
26 6 V
26 6 V
27 6 V
26 5 V
26 6 V
26 6 V
26 6 V
0.200 UL
LT6
593 1625 M
26 2 V
26 1 V
26 2 V
26 1 V
27 2 V
26 2 V
26 2 V
26 2 V
26 2 V
26 1 V
26 2 V
27 2 V
26 2 V
26 3 V
26 2 V
26 2 V
26 2 V
26 2 V
27 3 V
26 2 V
26 2 V
26 3 V
26 2 V
26 3 V
26 2 V
27 3 V
26 3 V
26 2 V
26 3 V
26 3 V
26 3 V
26 2 V
27 3 V
26 3 V
26 3 V
26 3 V
26 3 V
26 3 V
26 3 V
27 3 V
26 4 V
26 3 V
26 3 V
26 3 V
26 3 V
26 4 V
27 3 V
26 4 V
26 3 V
26 3 V
26 4 V
26 4 V
26 3 V
27 4 V
26 3 V
26 4 V
26 4 V
26 3 V
26 4 V
26 4 V
27 4 V
26 4 V
26 4 V
26 4 V
26 3 V
26 4 V
26 4 V
27 4 V
26 5 V
26 4 V
26 4 V
26 4 V
26 4 V
26 4 V
27 5 V
26 4 V
26 4 V
26 4 V
26 5 V
26 4 V
27 5 V
26 4 V
26 4 V
26 5 V
26 4 V
26 5 V
26 5 V
27 4 V
26 5 V
26 4 V
26 5 V
26 5 V
26 4 V
26 5 V
27 5 V
26 5 V
26 4 V
26 5 V
26 5 V
1.000 UL
LTb
350 300 M
3100 0 V
0 1760 V
-3100 0 V
350 300 L
1.000 UP
stroke
grestore
end
showpage
}}%
\put(3037,403){\makebox(0,0)[r]{\scriptsize $2\times$g.s,      P=-, $\beta=40.00$}}%
\put(3037,483){\makebox(0,0)[r]{\scriptsize $2\times 2^{nd}$, P=+, $\beta=40.00$}}%
\put(3037,563){\makebox(0,0)[r]{\scriptsize $1^{st}$, P=+, $\beta=40.00$}}%
\put(3037,643){\makebox(0,0)[r]{\scriptsize g.s,      P=+, $\beta=40.00$}}%
\put(3037,723){\makebox(0,0)[r]{\scriptsize $2\times$g.s,      P=-, $\beta=21.00$}}%
\put(3037,803){\makebox(0,0)[r]{\scriptsize $2\times 2^{nd}$, P=+, $\beta=21.00$}}%
\put(3037,883){\makebox(0,0)[r]{\scriptsize $1^{st}$, P=+,  $\beta=21.00$}}%
\put(3037,963){\makebox(0,0)[r]{\scriptsize g.s,      P=+, $\beta=21.00$}}%
\put(1900,50){\makebox(0,0){$l \sqrt{\sigma}$}}%
\put(100,1180){%
\special{ps: gsave currentpoint currentpoint translate
0 rotate neg exch neg exch translate}%
\makebox(0,0)[b]{\shortstack{$E/ \sqrt{\sigma}$}}%
\special{ps: currentpoint grestore moveto}%
}%
\put(3450,200){\makebox(0,0){ 6}}%
\put(3140,200){\makebox(0,0){ 5.5}}%
\put(2830,200){\makebox(0,0){ 5}}%
\put(2520,200){\makebox(0,0){ 4.5}}%
\put(2210,200){\makebox(0,0){ 4}}%
\put(1900,200){\makebox(0,0){ 3.5}}%
\put(1590,200){\makebox(0,0){ 3}}%
\put(1280,200){\makebox(0,0){ 2.5}}%
\put(970,200){\makebox(0,0){ 2}}%
\put(660,200){\makebox(0,0){ 1.5}}%
\put(350,200){\makebox(0,0){ 1}}%
\put(300,1967){\makebox(0,0)[r]{ 9}}%
\put(300,1782){\makebox(0,0)[r]{ 8}}%
\put(300,1597){\makebox(0,0)[r]{ 7}}%
\put(300,1412){\makebox(0,0)[r]{ 6}}%
\put(300,1226){\makebox(0,0)[r]{ 5}}%
\put(300,1041){\makebox(0,0)[r]{ 4}}%
\put(300,856){\makebox(0,0)[r]{ 3}}%
\put(300,671){\makebox(0,0)[r]{ 2}}%
\put(300,485){\makebox(0,0)[r]{ 1}}%
\put(300,300){\makebox(0,0)[r]{ 0}}%
\end{picture}%
\endgroup
 

%% file: plotsu62.tex
\begingroup%
  \makeatletter%
  \newcommand{\GNUPLOTspecial}{%
    \@sanitize\catcode`\%=14\relax\special}%
  \setlength{\unitlength}{0.1bp}%
\begin{picture}(3600,2160)(0,0)%
{\GNUPLOTspecial{"
/gnudict 256 dict def
gnudict begin
/Color true def
/Solid true def
/gnulinewidth 5.000 def
/userlinewidth gnulinewidth def
/vshift -33 def
/dl {10.0 mul} def
/hpt_ 31.5 def
/vpt_ 31.5 def
/hpt hpt_ def
/vpt vpt_ def
/Rounded false def
/M {moveto} bind def
/L {lineto} bind def
/R {rmoveto} bind def
/V {rlineto} bind def
/N {newpath moveto} bind def
/C {setrgbcolor} bind def
/f {rlineto fill} bind def
/vpt2 vpt 2 mul def
/hpt2 hpt 2 mul def
/Lshow { currentpoint stroke M
  0 vshift R show } def
/Rshow { currentpoint stroke M
  dup stringwidth pop neg vshift R show } def
/Cshow { currentpoint stroke M
  dup stringwidth pop -2 div vshift R show } def
/UP { dup vpt_ mul /vpt exch def hpt_ mul /hpt exch def
  /hpt2 hpt 2 mul def /vpt2 vpt 2 mul def } def
/DL { Color {setrgbcolor Solid {pop []} if 0 setdash }
 {pop pop pop 0 setgray Solid {pop []} if 0 setdash} ifelse } def
/BL { stroke userlinewidth 2 mul setlinewidth
      Rounded { 1 setlinejoin 1 setlinecap } if } def
/AL { stroke userlinewidth 2 div setlinewidth
      Rounded { 1 setlinejoin 1 setlinecap } if } def
/UL { dup gnulinewidth mul /userlinewidth exch def
      dup 1 lt {pop 1} if 10 mul /udl exch def } def
/PL { stroke userlinewidth setlinewidth
      Rounded { 1 setlinejoin 1 setlinecap } if } def
/LTw { PL [] 1 setgray } def
/LTb { BL [] 0 0 0 DL } def
/LTa { AL [1 udl mul 2 udl mul] 0 setdash 0 0 0 setrgbcolor } def
/LT0 { PL [] 1 0 0 DL } def
/LT1 { PL [4 dl 2 dl] 0 1 0 DL } def
/LT2 { PL [2 dl 3 dl] 0 0 1 DL } def
/LT3 { PL [1 dl 1.5 dl] 1 0 1 DL } def
/LT4 { PL [5 dl 2 dl 1 dl 2 dl] 0 1 1 DL } def
/LT5 { PL [4 dl 3 dl 1 dl 3 dl] 1 1 0 DL } def
/LT6 { PL [2 dl 2 dl 2 dl 4 dl] 0 0 0 DL } def
/LT7 { PL [2 dl 2 dl 2 dl 2 dl 2 dl 4 dl] 1 0.3 0 DL } def
/LT8 { PL [2 dl 2 dl 2 dl 2 dl 2 dl 2 dl 2 dl 4 dl] 0.5 0.5 0.5 DL } def
/Pnt { stroke [] 0 setdash
   gsave 1 setlinecap M 0 0 V stroke grestore } def
/Dia { stroke [] 0 setdash 2 copy vpt add M
  hpt neg vpt neg V hpt vpt neg V
  hpt vpt V hpt neg vpt V closepath stroke
  Pnt } def
/Pls { stroke [] 0 setdash vpt sub M 0 vpt2 V
  currentpoint stroke M
  hpt neg vpt neg R hpt2 0 V stroke
  } def
/Box { stroke [] 0 setdash 2 copy exch hpt sub exch vpt add M
  0 vpt2 neg V hpt2 0 V 0 vpt2 V
  hpt2 neg 0 V closepath stroke
  Pnt } def
/Crs { stroke [] 0 setdash exch hpt sub exch vpt add M
  hpt2 vpt2 neg V currentpoint stroke M
  hpt2 neg 0 R hpt2 vpt2 V stroke } def
/TriU { stroke [] 0 setdash 2 copy vpt 1.12 mul add M
  hpt neg vpt -1.62 mul V
  hpt 2 mul 0 V
  hpt neg vpt 1.62 mul V closepath stroke
  Pnt  } def
/Star { 2 copy Pls Crs } def
/BoxF { stroke [] 0 setdash exch hpt sub exch vpt add M
  0 vpt2 neg V  hpt2 0 V  0 vpt2 V
  hpt2 neg 0 V  closepath fill } def
/TriUF { stroke [] 0 setdash vpt 1.12 mul add M
  hpt neg vpt -1.62 mul V
  hpt 2 mul 0 V
  hpt neg vpt 1.62 mul V closepath fill } def
/TriD { stroke [] 0 setdash 2 copy vpt 1.12 mul sub M
  hpt neg vpt 1.62 mul V
  hpt 2 mul 0 V
  hpt neg vpt -1.62 mul V closepath stroke
  Pnt  } def
/TriDF { stroke [] 0 setdash vpt 1.12 mul sub M
  hpt neg vpt 1.62 mul V
  hpt 2 mul 0 V
  hpt neg vpt -1.62 mul V closepath fill} def
/DiaF { stroke [] 0 setdash vpt add M
  hpt neg vpt neg V hpt vpt neg V
  hpt vpt V hpt neg vpt V closepath fill } def
/Pent { stroke [] 0 setdash 2 copy gsave
  translate 0 hpt M 4 {72 rotate 0 hpt L} repeat
  closepath stroke grestore Pnt } def
/PentF { stroke [] 0 setdash gsave
  translate 0 hpt M 4 {72 rotate 0 hpt L} repeat
  closepath fill grestore } def
/Circle { stroke [] 0 setdash 2 copy
  hpt 0 360 arc stroke Pnt } def
/CircleF { stroke [] 0 setdash hpt 0 360 arc fill } def
/C0 { BL [] 0 setdash 2 copy moveto vpt 90 450  arc } bind def
/C1 { BL [] 0 setdash 2 copy        moveto
       2 copy  vpt 0 90 arc closepath fill
               vpt 0 360 arc closepath } bind def
/C2 { BL [] 0 setdash 2 copy moveto
       2 copy  vpt 90 180 arc closepath fill
               vpt 0 360 arc closepath } bind def
/C3 { BL [] 0 setdash 2 copy moveto
       2 copy  vpt 0 180 arc closepath fill
               vpt 0 360 arc closepath } bind def
/C4 { BL [] 0 setdash 2 copy moveto
       2 copy  vpt 180 270 arc closepath fill
               vpt 0 360 arc closepath } bind def
/C5 { BL [] 0 setdash 2 copy moveto
       2 copy  vpt 0 90 arc
       2 copy moveto
       2 copy  vpt 180 270 arc closepath fill
               vpt 0 360 arc } bind def
/C6 { BL [] 0 setdash 2 copy moveto
      2 copy  vpt 90 270 arc closepath fill
              vpt 0 360 arc closepath } bind def
/C7 { BL [] 0 setdash 2 copy moveto
      2 copy  vpt 0 270 arc closepath fill
              vpt 0 360 arc closepath } bind def
/C8 { BL [] 0 setdash 2 copy moveto
      2 copy vpt 270 360 arc closepath fill
              vpt 0 360 arc closepath } bind def
/C9 { BL [] 0 setdash 2 copy moveto
      2 copy  vpt 270 450 arc closepath fill
              vpt 0 360 arc closepath } bind def
/C10 { BL [] 0 setdash 2 copy 2 copy moveto vpt 270 360 arc closepath fill
       2 copy moveto
       2 copy vpt 90 180 arc closepath fill
               vpt 0 360 arc closepath } bind def
/C11 { BL [] 0 setdash 2 copy moveto
       2 copy  vpt 0 180 arc closepath fill
       2 copy moveto
       2 copy  vpt 270 360 arc closepath fill
               vpt 0 360 arc closepath } bind def
/C12 { BL [] 0 setdash 2 copy moveto
       2 copy  vpt 180 360 arc closepath fill
               vpt 0 360 arc closepath } bind def
/C13 { BL [] 0 setdash  2 copy moveto
       2 copy  vpt 0 90 arc closepath fill
       2 copy moveto
       2 copy  vpt 180 360 arc closepath fill
               vpt 0 360 arc closepath } bind def
/C14 { BL [] 0 setdash 2 copy moveto
       2 copy  vpt 90 360 arc closepath fill
               vpt 0 360 arc } bind def
/C15 { BL [] 0 setdash 2 copy vpt 0 360 arc closepath fill
               vpt 0 360 arc closepath } bind def
/Rec   { newpath 4 2 roll moveto 1 index 0 rlineto 0 exch rlineto
       neg 0 rlineto closepath } bind def
/Square { dup Rec } bind def
/Bsquare { vpt sub exch vpt sub exch vpt2 Square } bind def
/S0 { BL [] 0 setdash 2 copy moveto 0 vpt rlineto BL Bsquare } bind def
/S1 { BL [] 0 setdash 2 copy vpt Square fill Bsquare } bind def
/S2 { BL [] 0 setdash 2 copy exch vpt sub exch vpt Square fill Bsquare } bind def
/S3 { BL [] 0 setdash 2 copy exch vpt sub exch vpt2 vpt Rec fill Bsquare } bind def
/S4 { BL [] 0 setdash 2 copy exch vpt sub exch vpt sub vpt Square fill Bsquare } bind def
/S5 { BL [] 0 setdash 2 copy 2 copy vpt Square fill
       exch vpt sub exch vpt sub vpt Square fill Bsquare } bind def
/S6 { BL [] 0 setdash 2 copy exch vpt sub exch vpt sub vpt vpt2 Rec fill Bsquare } bind def
/S7 { BL [] 0 setdash 2 copy exch vpt sub exch vpt sub vpt vpt2 Rec fill
       2 copy vpt Square fill
       Bsquare } bind def
/S8 { BL [] 0 setdash 2 copy vpt sub vpt Square fill Bsquare } bind def
/S9 { BL [] 0 setdash 2 copy vpt sub vpt vpt2 Rec fill Bsquare } bind def
/S10 { BL [] 0 setdash 2 copy vpt sub vpt Square fill 2 copy exch vpt sub exch vpt Square fill
       Bsquare } bind def
/S11 { BL [] 0 setdash 2 copy vpt sub vpt Square fill 2 copy exch vpt sub exch vpt2 vpt Rec fill
       Bsquare } bind def
/S12 { BL [] 0 setdash 2 copy exch vpt sub exch vpt sub vpt2 vpt Rec fill Bsquare } bind def
/S13 { BL [] 0 setdash 2 copy exch vpt sub exch vpt sub vpt2 vpt Rec fill
       2 copy vpt Square fill Bsquare } bind def
/S14 { BL [] 0 setdash 2 copy exch vpt sub exch vpt sub vpt2 vpt Rec fill
       2 copy exch vpt sub exch vpt Square fill Bsquare } bind def
/S15 { BL [] 0 setdash 2 copy Bsquare fill Bsquare } bind def
/D0 { gsave translate 45 rotate 0 0 S0 stroke grestore } bind def
/D1 { gsave translate 45 rotate 0 0 S1 stroke grestore } bind def
/D2 { gsave translate 45 rotate 0 0 S2 stroke grestore } bind def
/D3 { gsave translate 45 rotate 0 0 S3 stroke grestore } bind def
/D4 { gsave translate 45 rotate 0 0 S4 stroke grestore } bind def
/D5 { gsave translate 45 rotate 0 0 S5 stroke grestore } bind def
/D6 { gsave translate 45 rotate 0 0 S6 stroke grestore } bind def
/D7 { gsave translate 45 rotate 0 0 S7 stroke grestore } bind def
/D8 { gsave translate 45 rotate 0 0 S8 stroke grestore } bind def
/D9 { gsave translate 45 rotate 0 0 S9 stroke grestore } bind def
/D10 { gsave translate 45 rotate 0 0 S10 stroke grestore } bind def
/D11 { gsave translate 45 rotate 0 0 S11 stroke grestore } bind def
/D12 { gsave translate 45 rotate 0 0 S12 stroke grestore } bind def
/D13 { gsave translate 45 rotate 0 0 S13 stroke grestore } bind def
/D14 { gsave translate 45 rotate 0 0 S14 stroke grestore } bind def
/D15 { gsave translate 45 rotate 0 0 S15 stroke grestore } bind def
/DiaE { stroke [] 0 setdash vpt add M
  hpt neg vpt neg V hpt vpt neg V
  hpt vpt V hpt neg vpt V closepath stroke } def
/BoxE { stroke [] 0 setdash exch hpt sub exch vpt add M
  0 vpt2 neg V hpt2 0 V 0 vpt2 V
  hpt2 neg 0 V closepath stroke } def
/TriUE { stroke [] 0 setdash vpt 1.12 mul add M
  hpt neg vpt -1.62 mul V
  hpt 2 mul 0 V
  hpt neg vpt 1.62 mul V closepath stroke } def
/TriDE { stroke [] 0 setdash vpt 1.12 mul sub M
  hpt neg vpt 1.62 mul V
  hpt 2 mul 0 V
  hpt neg vpt -1.62 mul V closepath stroke } def
/PentE { stroke [] 0 setdash gsave
  translate 0 hpt M 4 {72 rotate 0 hpt L} repeat
  closepath stroke grestore } def
/CircE { stroke [] 0 setdash 
  hpt 0 360 arc stroke } def
/Opaque { gsave closepath 1 setgray fill grestore 0 setgray closepath } def
/DiaW { stroke [] 0 setdash vpt add M
  hpt neg vpt neg V hpt vpt neg V
  hpt vpt V hpt neg vpt V Opaque stroke } def
/BoxW { stroke [] 0 setdash exch hpt sub exch vpt add M
  0 vpt2 neg V hpt2 0 V 0 vpt2 V
  hpt2 neg 0 V Opaque stroke } def
/TriUW { stroke [] 0 setdash vpt 1.12 mul add M
  hpt neg vpt -1.62 mul V
  hpt 2 mul 0 V
  hpt neg vpt 1.62 mul V Opaque stroke } def
/TriDW { stroke [] 0 setdash vpt 1.12 mul sub M
  hpt neg vpt 1.62 mul V
  hpt 2 mul 0 V
  hpt neg vpt -1.62 mul V Opaque stroke } def
/PentW { stroke [] 0 setdash gsave
  translate 0 hpt M 4 {72 rotate 0 hpt L} repeat
  Opaque stroke grestore } def
/CircW { stroke [] 0 setdash 
  hpt 0 360 arc Opaque stroke } def
/BoxFill { gsave Rec 1 setgray fill grestore } def
/BoxColFill {
  gsave Rec
  /Fillden exch def
  currentrgbcolor
  /ColB exch def /ColG exch def /ColR exch def
  /ColR ColR Fillden mul Fillden sub 1 add def
  /ColG ColG Fillden mul Fillden sub 1 add def
  /ColB ColB Fillden mul Fillden sub 1 add def
  ColR ColG ColB setrgbcolor
  fill grestore } def
%
%
/PatternFill { gsave /PFa [ 9 2 roll ] def
    PFa 0 get PFa 2 get 2 div add PFa 1 get PFa 3 get 2 div add translate
    PFa 2 get -2 div PFa 3 get -2 div PFa 2 get PFa 3 get Rec
    gsave 1 setgray fill grestore clip
    currentlinewidth 0.5 mul setlinewidth
    /PFs PFa 2 get dup mul PFa 3 get dup mul add sqrt def
    0 0 M PFa 5 get rotate PFs -2 div dup translate
	0 1 PFs PFa 4 get div 1 add floor cvi
	{ PFa 4 get mul 0 M 0 PFs V } for
    0 PFa 6 get ne {
	0 1 PFs PFa 4 get div 1 add floor cvi
	{ PFa 4 get mul 0 2 1 roll M PFs 0 V } for
    } if
    stroke grestore } def
/Symbol-Oblique /Symbol findfont [1 0 .167 1 0 0] makefont
dup length dict begin {1 index /FID eq {pop pop} {def} ifelse} forall
currentdict end definefont pop
end
gnudict begin
gsave
0 0 translate
0.100 0.100 scale
0 setgray
newpath
1.000 UL
LTb
350 300 M
63 0 V
3037 0 R
-63 0 V
1.000 UL
LTb
350 520 M
63 0 V
3037 0 R
-63 0 V
1.000 UL
LTb
350 740 M
63 0 V
3037 0 R
-63 0 V
1.000 UL
LTb
350 960 M
63 0 V
3037 0 R
-63 0 V
1.000 UL
LTb
350 1180 M
63 0 V
3037 0 R
-63 0 V
1.000 UL
LTb
350 1400 M
63 0 V
3037 0 R
-63 0 V
1.000 UL
LTb
350 1620 M
63 0 V
3037 0 R
-63 0 V
1.000 UL
LTb
350 1840 M
63 0 V
3037 0 R
-63 0 V
1.000 UL
LTb
350 2060 M
63 0 V
3037 0 R
-63 0 V
1.000 UL
LTb
350 300 M
0 63 V
0 1697 R
0 -63 V
1.000 UL
LTb
970 300 M
0 63 V
0 1697 R
0 -63 V
1.000 UL
LTb
1590 300 M
0 63 V
0 1697 R
0 -63 V
1.000 UL
LTb
2210 300 M
0 63 V
0 1697 R
0 -63 V
1.000 UL
LTb
2830 300 M
0 63 V
0 1697 R
0 -63 V
1.000 UL
LTb
3450 300 M
0 63 V
0 1697 R
0 -63 V
1.000 UL
LTb
1.000 UL
LTb
350 300 M
3100 0 V
0 1760 V
-3100 0 V
350 300 L
LTb
LTb
1.000 UP
0.800 UP
0.500 UL
LT0
LTb
LT0
3087 713 M
263 0 V
-263 31 R
0 -62 V
263 62 R
0 -62 V
428 472 M
0 3 V
-31 -3 R
62 0 V
-62 3 R
62 0 V
396 82 R
0 3 V
-31 -3 R
62 0 V
-62 3 R
62 0 V
395 81 R
0 3 V
-31 -3 R
62 0 V
-62 3 R
62 0 V
395 77 R
0 3 V
-31 -3 R
62 0 V
-62 3 R
62 0 V
396 78 R
0 4 V
-31 -4 R
62 0 V
-62 4 R
62 0 V
395 71 R
0 4 V
-31 -4 R
62 0 V
-62 4 R
62 0 V
396 76 R
0 5 V
-31 -5 R
62 0 V
-62 5 R
62 0 V
428 474 Pls
855 559 Pls
1281 643 Pls
1707 722 Pls
2134 804 Pls
2560 879 Pls
2987 959 Pls
3218 713 Pls
0.800 UP
0.500 UL
LT2
LTb
LT2
3087 613 M
263 0 V
-263 31 R
0 -62 V
263 62 R
0 -62 V
428 1199 M
0 9 V
-31 -9 R
62 0 V
-62 9 R
62 0 V
396 30 R
0 8 V
-31 -8 R
62 0 V
-62 8 R
62 0 V
395 36 R
0 10 V
-31 -10 R
62 0 V
-62 10 R
62 0 V
395 37 R
0 8 V
-31 -8 R
62 0 V
-62 8 R
62 0 V
396 51 R
0 11 V
-31 -11 R
62 0 V
-62 11 R
62 0 V
395 34 R
0 11 V
-31 -11 R
62 0 V
-62 11 R
62 0 V
396 42 R
0 10 V
-31 -10 R
62 0 V
-62 10 R
62 0 V
428 1204 Crs
855 1242 Crs
1281 1287 Crs
1707 1333 Crs
2134 1393 Crs
2560 1438 Crs
2987 1491 Crs
3218 613 Crs
0.800 UP
0.500 UL
LT0
LTb
LT0
3087 513 M
263 0 V
-263 31 R
0 -62 V
263 62 R
0 -62 V
428 1615 M
0 16 V
-31 -16 R
62 0 V
-62 16 R
62 0 V
-31 53 R
0 15 V
-31 -15 R
62 0 V
-62 15 R
62 0 V
396 -7 R
0 17 V
-31 -17 R
62 0 V
-62 17 R
62 0 V
-31 -94 R
0 18 V
-31 -18 R
62 0 V
-62 18 R
62 0 V
395 14 R
0 20 V
-31 -20 R
62 0 V
-62 20 R
62 0 V
-31 34 R
0 19 V
-31 -19 R
62 0 V
-62 19 R
62 0 V
395 -18 R
0 49 V
-31 -49 R
62 0 V
-62 49 R
62 0 V
-31 -20 R
0 17 V
-31 -17 R
62 0 V
-62 17 R
62 0 V
396 -37 R
0 14 V
-31 -14 R
62 0 V
-62 14 R
62 0 V
-31 40 R
0 18 V
-31 -18 R
62 0 V
-62 18 R
62 0 V
395 -22 R
0 18 V
-31 -18 R
62 0 V
-62 18 R
62 0 V
-31 46 R
0 19 V
-31 -19 R
62 0 V
-62 19 R
62 0 V
396 -33 R
0 20 V
-31 -20 R
62 0 V
-62 20 R
62 0 V
-31 17 R
0 18 V
-31 -18 R
62 0 V
-62 18 R
62 0 V
428 1623 Star
428 1691 Star
855 1701 Star
855 1624 Star
1281 1657 Star
1281 1710 Star
1707 1727 Star
1707 1740 Star
2134 1718 Star
2134 1774 Star
2560 1770 Star
2560 1835 Star
2987 1821 Star
2987 1857 Star
3218 513 Star
0.800 UP
0.500 UL
LT2
LTb
LT2
3087 413 M
263 0 V
-263 31 R
0 -62 V
263 62 R
0 -62 V
428 1546 M
0 44 V
-31 -44 R
62 0 V
-62 44 R
62 0 V
-31 127 R
0 66 V
-31 -66 R
62 0 V
-62 66 R
62 0 V
396 -78 R
0 17 V
-31 -17 R
62 0 V
-62 17 R
62 0 V
-31 77 R
0 24 V
-31 -24 R
62 0 V
-62 24 R
62 0 V
395 -119 R
0 15 V
-31 -15 R
62 0 V
-62 15 R
62 0 V
-31 87 R
0 17 V
-31 -17 R
62 0 V
-62 17 R
62 0 V
395 -108 R
0 15 V
-31 -15 R
62 0 V
-62 15 R
62 0 V
-31 42 R
0 17 V
-31 -17 R
62 0 V
-62 17 R
62 0 V
396 -21 R
0 18 V
-31 -18 R
62 0 V
-62 18 R
62 0 V
-31 21 R
0 20 V
-31 -20 R
62 0 V
-62 20 R
62 0 V
395 -17 R
0 24 V
-31 -24 R
62 0 V
-62 24 R
62 0 V
-31 7 R
0 19 V
-31 -19 R
62 0 V
-62 19 R
62 0 V
396 -3 R
0 19 V
-31 -19 R
62 0 V
-62 19 R
62 0 V
-31 4 R
0 21 V
-31 -21 R
62 0 V
-62 21 R
62 0 V
428 1568 Box
428 1750 Box
855 1713 Box
855 1811 Box
1281 1712 Box
1281 1815 Box
1707 1722 Box
1707 1780 Box
2134 1777 Box
2134 1817 Box
2560 1822 Box
2560 1850 Box
2987 1866 Box
2987 1890 Box
3218 413 Box
0.500 UL
LT0
428 474 M
26 5 V
26 6 V
26 5 V
26 5 V
25 5 V
26 6 V
26 5 V
26 5 V
26 5 V
26 5 V
26 5 V
25 5 V
26 6 V
26 5 V
26 5 V
26 5 V
26 5 V
25 5 V
26 5 V
26 5 V
26 5 V
26 5 V
26 5 V
25 5 V
26 5 V
26 5 V
26 5 V
26 5 V
26 5 V
26 5 V
25 5 V
26 5 V
26 5 V
26 5 V
26 5 V
26 5 V
25 4 V
26 5 V
26 5 V
26 5 V
26 5 V
26 5 V
25 5 V
26 5 V
26 5 V
26 5 V
26 4 V
26 5 V
25 5 V
26 5 V
26 5 V
26 5 V
26 5 V
26 4 V
26 5 V
25 5 V
26 5 V
26 5 V
26 5 V
26 4 V
26 5 V
25 5 V
26 5 V
26 5 V
26 4 V
26 5 V
26 5 V
25 5 V
26 5 V
26 4 V
26 5 V
26 5 V
26 5 V
26 5 V
25 4 V
26 5 V
26 5 V
26 5 V
26 4 V
26 5 V
25 5 V
26 5 V
26 4 V
26 5 V
26 5 V
26 5 V
25 5 V
26 4 V
26 5 V
26 5 V
26 5 V
26 4 V
26 5 V
25 5 V
26 4 V
26 5 V
26 5 V
26 5 V
26 4 V
0.500 UL
LT2
428 1251 M
26 2 V
26 2 V
26 2 V
26 1 V
25 2 V
26 2 V
26 2 V
26 2 V
26 2 V
26 2 V
26 2 V
25 2 V
26 2 V
26 2 V
26 2 V
26 2 V
26 2 V
25 2 V
26 2 V
26 2 V
26 2 V
26 2 V
26 2 V
25 2 V
26 2 V
26 2 V
26 3 V
26 2 V
26 2 V
26 2 V
25 2 V
26 2 V
26 3 V
26 2 V
26 2 V
26 3 V
25 2 V
26 2 V
26 2 V
26 3 V
26 2 V
26 2 V
25 3 V
26 2 V
26 3 V
26 2 V
26 2 V
26 3 V
25 2 V
26 3 V
26 2 V
26 3 V
26 2 V
26 3 V
26 2 V
25 3 V
26 2 V
26 3 V
26 2 V
26 3 V
26 2 V
25 3 V
26 2 V
26 3 V
26 3 V
26 2 V
26 3 V
25 3 V
26 2 V
26 3 V
26 3 V
26 2 V
26 3 V
26 3 V
25 2 V
26 3 V
26 3 V
26 3 V
26 2 V
26 3 V
25 3 V
26 3 V
26 3 V
26 2 V
26 3 V
26 3 V
25 3 V
26 3 V
26 3 V
26 2 V
26 3 V
26 3 V
26 3 V
25 3 V
26 3 V
26 3 V
26 3 V
26 3 V
26 3 V
0.500 UL
LTb
428 1689 M
26 1 V
26 1 V
26 2 V
26 1 V
25 1 V
26 2 V
26 1 V
26 2 V
26 1 V
26 1 V
26 2 V
25 1 V
26 2 V
26 1 V
26 2 V
26 1 V
26 1 V
25 2 V
26 1 V
26 2 V
26 2 V
26 1 V
26 2 V
25 1 V
26 2 V
26 1 V
26 2 V
26 2 V
26 1 V
26 2 V
25 2 V
26 1 V
26 2 V
26 2 V
26 1 V
26 2 V
25 2 V
26 1 V
26 2 V
26 2 V
26 2 V
26 1 V
25 2 V
26 2 V
26 2 V
26 2 V
26 1 V
26 2 V
25 2 V
26 2 V
26 2 V
26 2 V
26 1 V
26 2 V
26 2 V
25 2 V
26 2 V
26 2 V
26 2 V
26 2 V
26 2 V
25 2 V
26 2 V
26 2 V
26 2 V
26 2 V
26 2 V
25 2 V
26 2 V
26 2 V
26 2 V
26 2 V
26 2 V
26 2 V
25 2 V
26 3 V
26 2 V
26 2 V
26 2 V
26 2 V
25 2 V
26 2 V
26 3 V
26 2 V
26 2 V
26 2 V
25 2 V
26 3 V
26 2 V
26 2 V
26 2 V
26 3 V
26 2 V
25 2 V
26 2 V
26 3 V
26 2 V
26 2 V
26 3 V
1.000 UL
LTb
350 300 M
3100 0 V
0 1760 V
-3100 0 V
350 300 L
1.000 UP
stroke
grestore
end
showpage
}}%
\put(3037,413){\makebox(0,0)[r]{$2\times$g.s,      P=-, $\beta=90.00$}}%
\put(3037,513){\makebox(0,0)[r]{$2\times 2^{nd}$, P=+, $\beta=90.00$}}%
\put(3037,613){\makebox(0,0)[r]{$1^{st}$, P=+, $\beta=90.00$}}%
\put(3037,713){\makebox(0,0)[r]{g.s,      P=+,  $\beta=90.00$}}%
\put(1900,50){\makebox(0,0){$l \sqrt{\sigma}$}}%
\put(100,1180){%
\special{ps: gsave currentpoint currentpoint translate
0 rotate neg exch neg exch translate}%
\makebox(0,0)[b]{\shortstack{$E/ \sqrt{\sigma}$}}%
\special{ps: currentpoint grestore moveto}%
}%
\put(3450,200){\makebox(0,0){ 4.5}}%
\put(2830,200){\makebox(0,0){ 4}}%
\put(2210,200){\makebox(0,0){ 3.5}}%
\put(1590,200){\makebox(0,0){ 3}}%
\put(970,200){\makebox(0,0){ 2.5}}%
\put(350,200){\makebox(0,0){ 2}}%
\put(300,2060){\makebox(0,0)[r]{ 9}}%
\put(300,1840){\makebox(0,0)[r]{ 8}}%
\put(300,1620){\makebox(0,0)[r]{ 7}}%
\put(300,1400){\makebox(0,0)[r]{ 6}}%
\put(300,1180){\makebox(0,0)[r]{ 5}}%
\put(300,960){\makebox(0,0)[r]{ 4}}%
\put(300,740){\makebox(0,0)[r]{ 3}}%
\put(300,520){\makebox(0,0)[r]{ 2}}%
\put(300,300){\makebox(0,0)[r]{ 1}}%
\end{picture}%
\endgroup
 

%% file: mom4.tex
\begingroup%
  \makeatletter%
  \newcommand{\GNUPLOTspecial}{%
    \@sanitize\catcode`\%=14\relax\special}%
  \setlength{\unitlength}{0.1bp}%
\begin{picture}(3600,2160)(0,0)%
{\GNUPLOTspecial{"
/gnudict 256 dict def
gnudict begin
/Color true def
/Solid true def
/gnulinewidth 5.000 def
/userlinewidth gnulinewidth def
/vshift -33 def
/dl {10.0 mul} def
/hpt_ 31.5 def
/vpt_ 31.5 def
/hpt hpt_ def
/vpt vpt_ def
/Rounded false def
/M {moveto} bind def
/L {lineto} bind def
/R {rmoveto} bind def
/V {rlineto} bind def
/N {newpath moveto} bind def
/C {setrgbcolor} bind def
/f {rlineto fill} bind def
/vpt2 vpt 2 mul def
/hpt2 hpt 2 mul def
/Lshow { currentpoint stroke M
  0 vshift R show } def
/Rshow { currentpoint stroke M
  dup stringwidth pop neg vshift R show } def
/Cshow { currentpoint stroke M
  dup stringwidth pop -2 div vshift R show } def
/UP { dup vpt_ mul /vpt exch def hpt_ mul /hpt exch def
  /hpt2 hpt 2 mul def /vpt2 vpt 2 mul def } def
/DL { Color {setrgbcolor Solid {pop []} if 0 setdash }
 {pop pop pop 0 setgray Solid {pop []} if 0 setdash} ifelse } def
/BL { stroke userlinewidth 2 mul setlinewidth
      Rounded { 1 setlinejoin 1 setlinecap } if } def
/AL { stroke userlinewidth 2 div setlinewidth
      Rounded { 1 setlinejoin 1 setlinecap } if } def
/UL { dup gnulinewidth mul /userlinewidth exch def
      dup 1 lt {pop 1} if 10 mul /udl exch def } def
/PL { stroke userlinewidth setlinewidth
      Rounded { 1 setlinejoin 1 setlinecap } if } def
/LTw { PL [] 1 setgray } def
/LTb { BL [] 0 0 0 DL } def
/LTa { AL [1 udl mul 2 udl mul] 0 setdash 0 0 0 setrgbcolor } def
/LT0 { PL [] 1 0 0 DL } def
/LT1 { PL [4 dl 2 dl] 0 1 0 DL } def
/LT2 { PL [2 dl 3 dl] 0 0 1 DL } def
/LT3 { PL [1 dl 1.5 dl] 1 0 1 DL } def
/LT4 { PL [5 dl 2 dl 1 dl 2 dl] 0 1 1 DL } def
/LT5 { PL [4 dl 3 dl 1 dl 3 dl] 0 0 1 DL } def
/LT6 { PL [2 dl 2 dl 2 dl 4 dl] 0 0 0 DL } def
/LT7 { PL [2 dl 2 dl 2 dl 2 dl 2 dl 4 dl] 1 0.3 0 DL } def
/LT8 { PL [2 dl 2 dl 2 dl 2 dl 2 dl 2 dl 2 dl 4 dl] 0.5 0.5 0.5 DL } def
/Pnt { stroke [] 0 setdash
   gsave 1 setlinecap M 0 0 V stroke grestore } def
/Dia { stroke [] 0 setdash 2 copy vpt add M
  hpt neg vpt neg V hpt vpt neg V
  hpt vpt V hpt neg vpt V closepath stroke
  Pnt } def
/Pls { stroke [] 0 setdash vpt sub M 0 vpt2 V
  currentpoint stroke M
  hpt neg vpt neg R hpt2 0 V stroke
  } def
/Box { stroke [] 0 setdash 2 copy exch hpt sub exch vpt add M
  0 vpt2 neg V hpt2 0 V 0 vpt2 V
  hpt2 neg 0 V closepath stroke
  Pnt } def
/Crs { stroke [] 0 setdash exch hpt sub exch vpt add M
  hpt2 vpt2 neg V currentpoint stroke M
  hpt2 neg 0 R hpt2 vpt2 V stroke } def
/TriU { stroke [] 0 setdash 2 copy vpt 1.12 mul add M
  hpt neg vpt -1.62 mul V
  hpt 2 mul 0 V
  hpt neg vpt 1.62 mul V closepath stroke
  Pnt  } def
/Star { 2 copy Pls Crs } def
/BoxF { stroke [] 0 setdash exch hpt sub exch vpt add M
  0 vpt2 neg V  hpt2 0 V  0 vpt2 V
  hpt2 neg 0 V  closepath fill } def
/TriUF { stroke [] 0 setdash vpt 1.12 mul add M
  hpt neg vpt -1.62 mul V
  hpt 2 mul 0 V
  hpt neg vpt 1.62 mul V closepath fill } def
/TriD { stroke [] 0 setdash 2 copy vpt 1.12 mul sub M
  hpt neg vpt 1.62 mul V
  hpt 2 mul 0 V
  hpt neg vpt -1.62 mul V closepath stroke
  Pnt  } def
/TriDF { stroke [] 0 setdash vpt 1.12 mul sub M
  hpt neg vpt 1.62 mul V
  hpt 2 mul 0 V
  hpt neg vpt -1.62 mul V closepath fill} def
/DiaF { stroke [] 0 setdash vpt add M
  hpt neg vpt neg V hpt vpt neg V
  hpt vpt V hpt neg vpt V closepath fill } def
/Pent { stroke [] 0 setdash 2 copy gsave
  translate 0 hpt M 4 {72 rotate 0 hpt L} repeat
  closepath stroke grestore Pnt } def
/PentF { stroke [] 0 setdash gsave
  translate 0 hpt M 4 {72 rotate 0 hpt L} repeat
  closepath fill grestore } def
/Circle { stroke [] 0 setdash 2 copy
  hpt 0 360 arc stroke Pnt } def
/CircleF { stroke [] 0 setdash hpt 0 360 arc fill } def
/C0 { BL [] 0 setdash 2 copy moveto vpt 90 450  arc } bind def
/C1 { BL [] 0 setdash 2 copy        moveto
       2 copy  vpt 0 90 arc closepath fill
               vpt 0 360 arc closepath } bind def
/C2 { BL [] 0 setdash 2 copy moveto
       2 copy  vpt 90 180 arc closepath fill
               vpt 0 360 arc closepath } bind def
/C3 { BL [] 0 setdash 2 copy moveto
       2 copy  vpt 0 180 arc closepath fill
               vpt 0 360 arc closepath } bind def
/C4 { BL [] 0 setdash 2 copy moveto
       2 copy  vpt 180 270 arc closepath fill
               vpt 0 360 arc closepath } bind def
/C5 { BL [] 0 setdash 2 copy moveto
       2 copy  vpt 0 90 arc
       2 copy moveto
       2 copy  vpt 180 270 arc closepath fill
               vpt 0 360 arc } bind def
/C6 { BL [] 0 setdash 2 copy moveto
      2 copy  vpt 90 270 arc closepath fill
              vpt 0 360 arc closepath } bind def
/C7 { BL [] 0 setdash 2 copy moveto
      2 copy  vpt 0 270 arc closepath fill
              vpt 0 360 arc closepath } bind def
/C8 { BL [] 0 setdash 2 copy moveto
      2 copy vpt 270 360 arc closepath fill
              vpt 0 360 arc closepath } bind def
/C9 { BL [] 0 setdash 2 copy moveto
      2 copy  vpt 270 450 arc closepath fill
              vpt 0 360 arc closepath } bind def
/C10 { BL [] 0 setdash 2 copy 2 copy moveto vpt 270 360 arc closepath fill
       2 copy moveto
       2 copy vpt 90 180 arc closepath fill
               vpt 0 360 arc closepath } bind def
/C11 { BL [] 0 setdash 2 copy moveto
       2 copy  vpt 0 180 arc closepath fill
       2 copy moveto
       2 copy  vpt 270 360 arc closepath fill
               vpt 0 360 arc closepath } bind def
/C12 { BL [] 0 setdash 2 copy moveto
       2 copy  vpt 180 360 arc closepath fill
               vpt 0 360 arc closepath } bind def
/C13 { BL [] 0 setdash  2 copy moveto
       2 copy  vpt 0 90 arc closepath fill
       2 copy moveto
       2 copy  vpt 180 360 arc closepath fill
               vpt 0 360 arc closepath } bind def
/C14 { BL [] 0 setdash 2 copy moveto
       2 copy  vpt 90 360 arc closepath fill
               vpt 0 360 arc } bind def
/C15 { BL [] 0 setdash 2 copy vpt 0 360 arc closepath fill
               vpt 0 360 arc closepath } bind def
/Rec   { newpath 4 2 roll moveto 1 index 0 rlineto 0 exch rlineto
       neg 0 rlineto closepath } bind def
/Square { dup Rec } bind def
/Bsquare { vpt sub exch vpt sub exch vpt2 Square } bind def
/S0 { BL [] 0 setdash 2 copy moveto 0 vpt rlineto BL Bsquare } bind def
/S1 { BL [] 0 setdash 2 copy vpt Square fill Bsquare } bind def
/S2 { BL [] 0 setdash 2 copy exch vpt sub exch vpt Square fill Bsquare } bind def
/S3 { BL [] 0 setdash 2 copy exch vpt sub exch vpt2 vpt Rec fill Bsquare } bind def
/S4 { BL [] 0 setdash 2 copy exch vpt sub exch vpt sub vpt Square fill Bsquare } bind def
/S5 { BL [] 0 setdash 2 copy 2 copy vpt Square fill
       exch vpt sub exch vpt sub vpt Square fill Bsquare } bind def
/S6 { BL [] 0 setdash 2 copy exch vpt sub exch vpt sub vpt vpt2 Rec fill Bsquare } bind def
/S7 { BL [] 0 setdash 2 copy exch vpt sub exch vpt sub vpt vpt2 Rec fill
       2 copy vpt Square fill
       Bsquare } bind def
/S8 { BL [] 0 setdash 2 copy vpt sub vpt Square fill Bsquare } bind def
/S9 { BL [] 0 setdash 2 copy vpt sub vpt vpt2 Rec fill Bsquare } bind def
/S10 { BL [] 0 setdash 2 copy vpt sub vpt Square fill 2 copy exch vpt sub exch vpt Square fill
       Bsquare } bind def
/S11 { BL [] 0 setdash 2 copy vpt sub vpt Square fill 2 copy exch vpt sub exch vpt2 vpt Rec fill
       Bsquare } bind def
/S12 { BL [] 0 setdash 2 copy exch vpt sub exch vpt sub vpt2 vpt Rec fill Bsquare } bind def
/S13 { BL [] 0 setdash 2 copy exch vpt sub exch vpt sub vpt2 vpt Rec fill
       2 copy vpt Square fill Bsquare } bind def
/S14 { BL [] 0 setdash 2 copy exch vpt sub exch vpt sub vpt2 vpt Rec fill
       2 copy exch vpt sub exch vpt Square fill Bsquare } bind def
/S15 { BL [] 0 setdash 2 copy Bsquare fill Bsquare } bind def
/D0 { gsave translate 45 rotate 0 0 S0 stroke grestore } bind def
/D1 { gsave translate 45 rotate 0 0 S1 stroke grestore } bind def
/D2 { gsave translate 45 rotate 0 0 S2 stroke grestore } bind def
/D3 { gsave translate 45 rotate 0 0 S3 stroke grestore } bind def
/D4 { gsave translate 45 rotate 0 0 S4 stroke grestore } bind def
/D5 { gsave translate 45 rotate 0 0 S5 stroke grestore } bind def
/D6 { gsave translate 45 rotate 0 0 S6 stroke grestore } bind def
/D7 { gsave translate 45 rotate 0 0 S7 stroke grestore } bind def
/D8 { gsave translate 45 rotate 0 0 S8 stroke grestore } bind def
/D9 { gsave translate 45 rotate 0 0 S9 stroke grestore } bind def
/D10 { gsave translate 45 rotate 0 0 S10 stroke grestore } bind def
/D11 { gsave translate 45 rotate 0 0 S11 stroke grestore } bind def
/D12 { gsave translate 45 rotate 0 0 S12 stroke grestore } bind def
/D13 { gsave translate 45 rotate 0 0 S13 stroke grestore } bind def
/D14 { gsave translate 45 rotate 0 0 S14 stroke grestore } bind def
/D15 { gsave translate 45 rotate 0 0 S15 stroke grestore } bind def
/DiaE { stroke [] 0 setdash vpt add M
  hpt neg vpt neg V hpt vpt neg V
  hpt vpt V hpt neg vpt V closepath stroke } def
/BoxE { stroke [] 0 setdash exch hpt sub exch vpt add M
  0 vpt2 neg V hpt2 0 V 0 vpt2 V
  hpt2 neg 0 V closepath stroke } def
/TriUE { stroke [] 0 setdash vpt 1.12 mul add M
  hpt neg vpt -1.62 mul V
  hpt 2 mul 0 V
  hpt neg vpt 1.62 mul V closepath stroke } def
/TriDE { stroke [] 0 setdash vpt 1.12 mul sub M
  hpt neg vpt 1.62 mul V
  hpt 2 mul 0 V
  hpt neg vpt -1.62 mul V closepath stroke } def
/PentE { stroke [] 0 setdash gsave
  translate 0 hpt M 4 {72 rotate 0 hpt L} repeat
  closepath stroke grestore } def
/CircE { stroke [] 0 setdash 
  hpt 0 360 arc stroke } def
/Opaque { gsave closepath 1 setgray fill grestore 0 setgray closepath } def
/DiaW { stroke [] 0 setdash vpt add M
  hpt neg vpt neg V hpt vpt neg V
  hpt vpt V hpt neg vpt V Opaque stroke } def
/BoxW { stroke [] 0 setdash exch hpt sub exch vpt add M
  0 vpt2 neg V hpt2 0 V 0 vpt2 V
  hpt2 neg 0 V Opaque stroke } def
/TriUW { stroke [] 0 setdash vpt 1.12 mul add M
  hpt neg vpt -1.62 mul V
  hpt 2 mul 0 V
  hpt neg vpt 1.62 mul V Opaque stroke } def
/TriDW { stroke [] 0 setdash vpt 1.12 mul sub M
  hpt neg vpt 1.62 mul V
  hpt 2 mul 0 V
  hpt neg vpt -1.62 mul V Opaque stroke } def
/PentW { stroke [] 0 setdash gsave
  translate 0 hpt M 4 {72 rotate 0 hpt L} repeat
  Opaque stroke grestore } def
/CircW { stroke [] 0 setdash 
  hpt 0 360 arc Opaque stroke } def
/BoxFill { gsave Rec 1 setgray fill grestore } def
/BoxColFill {
  gsave Rec
  /Fillden exch def
  currentrgbcolor
  /ColB exch def /ColG exch def /ColR exch def
  /ColR ColR Fillden mul Fillden sub 1 add def
  /ColG ColG Fillden mul Fillden sub 1 add def
  /ColB ColB Fillden mul Fillden sub 1 add def
  ColR ColG ColB setrgbcolor
  fill grestore } def
%
%
/PatternFill { gsave /PFa [ 9 2 roll ] def
    PFa 0 get PFa 2 get 2 div add PFa 1 get PFa 3 get 2 div add translate
    PFa 2 get -2 div PFa 3 get -2 div PFa 2 get PFa 3 get Rec
    gsave 1 setgray fill grestore clip
    currentlinewidth 0.5 mul setlinewidth
    /PFs PFa 2 get dup mul PFa 3 get dup mul add sqrt def
    0 0 M PFa 5 get rotate PFs -2 div dup translate
	0 1 PFs PFa 4 get div 1 add floor cvi
	{ PFa 4 get mul 0 M 0 PFs V } for
    0 PFa 6 get ne {
	0 1 PFs PFa 4 get div 1 add floor cvi
	{ PFa 4 get mul 0 2 1 roll M PFs 0 V } for
    } if
    stroke grestore } def
/Symbol-Oblique /Symbol findfont [1 0 .167 1 0 0] makefont
dup length dict begin {1 index /FID eq {pop pop} {def} ifelse} forall
currentdict end definefont pop
end
gnudict begin
gsave
0 0 translate
0.100 0.100 scale
0 setgray
newpath
1.000 UL
LTb
350 300 M
63 0 V
3037 0 R
-63 0 V
1.000 UL
LTb
350 496 M
63 0 V
3037 0 R
-63 0 V
1.000 UL
LTb
350 691 M
63 0 V
3037 0 R
-63 0 V
1.000 UL
LTb
350 887 M
63 0 V
3037 0 R
-63 0 V
1.000 UL
LTb
350 1082 M
63 0 V
3037 0 R
-63 0 V
1.000 UL
LTb
350 1278 M
63 0 V
3037 0 R
-63 0 V
1.000 UL
LTb
350 1473 M
63 0 V
3037 0 R
-63 0 V
1.000 UL
LTb
350 1669 M
63 0 V
3037 0 R
-63 0 V
1.000 UL
LTb
350 1864 M
63 0 V
3037 0 R
-63 0 V
1.000 UL
LTb
350 2060 M
63 0 V
3037 0 R
-63 0 V
1.000 UL
LTb
350 300 M
0 63 V
0 1697 R
0 -63 V
1.000 UL
LTb
970 300 M
0 63 V
0 1697 R
0 -63 V
1.000 UL
LTb
1590 300 M
0 63 V
0 1697 R
0 -63 V
1.000 UL
LTb
2210 300 M
0 63 V
0 1697 R
0 -63 V
1.000 UL
LTb
2830 300 M
0 63 V
0 1697 R
0 -63 V
1.000 UL
LTb
3450 300 M
0 63 V
0 1697 R
0 -63 V
1.000 UL
LTb
1.000 UL
LTb
350 300 M
3100 0 V
0 1760 V
-3100 0 V
350 300 L
LTb
LTb
1.000 UP
0.400 UP
0.500 UL
LT2
LTb
LT2
3087 1203 M
263 0 V
-263 31 R
0 -62 V
263 62 R
0 -62 V
458 1067 M
0 8 V
-31 -8 R
62 0 V
-62 8 R
62 0 V
1263 128 R
0 7 V
-31 -7 R
62 0 V
-62 7 R
62 0 V
1264 140 R
0 7 V
-31 -7 R
62 0 V
-62 7 R
62 0 V
458 1071 Pls
1752 1206 Pls
3047 1353 Pls
3218 1203 Pls
0.400 UP
0.500 UL
LT0
LTb
LT0
3087 1123 M
263 0 V
-263 31 R
0 -62 V
263 62 R
0 -62 V
458 1542 M
0 11 V
-31 -11 R
62 0 V
-62 11 R
62 0 V
1263 57 R
0 14 V
-31 -14 R
62 0 V
-62 14 R
62 0 V
1264 103 R
0 16 V
-31 -16 R
62 0 V
-62 16 R
62 0 V
458 1547 Crs
1752 1617 Crs
3047 1735 Crs
3218 1123 Crs
0.400 UP
0.500 UL
LT2
LTb
LT2
3087 1043 M
263 0 V
-263 31 R
0 -62 V
263 62 R
0 -62 V
458 1556 M
0 13 V
-31 -13 R
62 0 V
-62 13 R
62 0 V
1263 55 R
0 12 V
-31 -12 R
62 0 V
-62 12 R
62 0 V
1264 89 R
0 17 V
-31 -17 R
62 0 V
-62 17 R
62 0 V
458 1563 Star
1752 1630 Star
3047 1734 Star
3218 1043 Star
0.400 UP
0.500 UL
LT0
LTb
LT0
3087 963 M
263 0 V
-263 31 R
0 -62 V
263 62 R
0 -62 V
458 1310 M
0 31 V
-31 -31 R
62 0 V
-62 31 R
62 0 V
1263 84 R
0 17 V
-31 -17 R
62 0 V
-62 17 R
62 0 V
1264 118 R
0 15 V
-31 -15 R
62 0 V
-62 15 R
62 0 V
458 1326 Box
1752 1433 Box
3047 1568 Box
3218 963 Box
0.400 UP
0.500 UL
LT2
LTb
LT2
3087 883 M
263 0 V
-263 31 R
0 -62 V
263 62 R
0 -62 V
458 1226 M
0 27 V
-31 -27 R
62 0 V
-62 27 R
62 0 V
1263 160 R
0 15 V
-31 -15 R
62 0 V
-62 15 R
62 0 V
1264 123 R
0 13 V
-31 -13 R
62 0 V
-62 13 R
62 0 V
458 1240 BoxF
1752 1420 BoxF
3047 1557 BoxF
3218 883 BoxF
0.400 UP
0.500 UL
LT0
LTb
LT0
3087 803 M
263 0 V
-263 31 R
0 -62 V
263 62 R
0 -62 V
458 1737 M
0 38 V
-31 -38 R
62 0 V
-62 38 R
62 0 V
-31 -62 R
0 45 V
-31 -45 R
62 0 V
-62 45 R
62 0 V
1263 46 R
0 25 V
-31 -25 R
62 0 V
-62 25 R
62 0 V
-31 -23 R
0 34 V
-31 -34 R
62 0 V
-62 34 R
62 0 V
1264 75 R
0 23 V
-31 -23 R
62 0 V
-62 23 R
62 0 V
-31 0 R
0 20 V
-31 -20 R
62 0 V
-62 20 R
62 0 V
458 1756 Circle
458 1736 Circle
1752 1817 Circle
1752 1823 Circle
3047 1926 Circle
3047 1948 Circle
3218 803 Circle
0.400 UP
0.500 UL
LT2
LTb
LT2
3087 723 M
263 0 V
-263 31 R
0 -62 V
263 62 R
0 -62 V
458 1711 M
0 44 V
-31 -44 R
62 0 V
-62 44 R
62 0 V
1263 67 R
0 31 V
-31 -31 R
62 0 V
-62 31 R
62 0 V
1264 64 R
0 26 V
-31 -26 R
62 0 V
-62 26 R
62 0 V
458 1733 CircleF
1752 1838 CircleF
3047 1930 CircleF
3218 723 CircleF
0.500 UL
LTb
LTb
LTb
3087 643 M
263 0 V
350 1070 M
31 3 V
32 2 V
31 3 V
31 3 V
32 2 V
31 3 V
31 2 V
32 3 V
31 3 V
31 3 V
31 2 V
32 3 V
31 3 V
31 3 V
32 3 V
31 2 V
31 3 V
32 3 V
31 3 V
31 3 V
32 3 V
31 3 V
31 3 V
32 3 V
31 3 V
31 3 V
31 3 V
32 3 V
31 3 V
31 3 V
32 3 V
31 3 V
31 4 V
32 3 V
31 3 V
31 3 V
32 3 V
31 4 V
31 3 V
32 3 V
31 3 V
31 4 V
31 3 V
32 3 V
31 4 V
31 3 V
32 3 V
31 4 V
31 3 V
32 4 V
31 3 V
31 4 V
32 3 V
31 3 V
31 4 V
32 3 V
31 4 V
31 3 V
31 4 V
32 4 V
31 3 V
31 4 V
32 3 V
31 4 V
31 3 V
32 4 V
31 4 V
31 3 V
32 4 V
31 4 V
31 3 V
32 4 V
31 4 V
31 3 V
31 4 V
32 4 V
31 4 V
31 3 V
32 4 V
31 4 V
31 4 V
32 3 V
31 4 V
31 4 V
32 4 V
31 4 V
31 4 V
32 3 V
31 4 V
31 4 V
31 4 V
32 4 V
31 4 V
31 4 V
32 4 V
31 3 V
31 4 V
32 4 V
31 4 V
0.500 UL
LT1
LTb
LT1
3087 563 M
263 0 V
350 1547 M
31 1 V
32 2 V
31 2 V
31 1 V
32 2 V
31 1 V
31 2 V
32 2 V
31 1 V
31 2 V
31 2 V
32 2 V
31 1 V
31 2 V
32 2 V
31 2 V
31 2 V
32 2 V
31 1 V
31 2 V
32 2 V
31 2 V
31 2 V
32 2 V
31 2 V
31 2 V
31 2 V
32 2 V
31 2 V
31 2 V
32 2 V
31 2 V
31 2 V
32 2 V
31 2 V
31 2 V
32 3 V
31 2 V
31 2 V
32 2 V
31 2 V
31 3 V
31 2 V
32 2 V
31 2 V
31 3 V
32 2 V
31 2 V
31 3 V
32 2 V
31 2 V
31 3 V
32 2 V
31 2 V
31 3 V
32 2 V
31 3 V
31 2 V
31 3 V
32 2 V
31 2 V
31 3 V
32 3 V
31 2 V
31 3 V
32 2 V
31 3 V
31 2 V
32 3 V
31 2 V
31 3 V
32 3 V
31 2 V
31 3 V
31 3 V
32 2 V
31 3 V
31 3 V
32 2 V
31 3 V
31 3 V
32 3 V
31 2 V
31 3 V
32 3 V
31 3 V
31 3 V
32 2 V
31 3 V
31 3 V
31 3 V
32 3 V
31 3 V
31 3 V
32 2 V
31 3 V
31 3 V
32 3 V
31 3 V
0.500 UL
LT3
LTb
LT3
3087 483 M
263 0 V
350 1336 M
31 2 V
32 2 V
31 2 V
31 2 V
32 2 V
31 2 V
31 2 V
32 2 V
31 2 V
31 2 V
31 2 V
32 2 V
31 2 V
31 2 V
32 3 V
31 2 V
31 2 V
32 2 V
31 2 V
31 3 V
32 2 V
31 2 V
31 2 V
32 3 V
31 2 V
31 2 V
31 3 V
32 2 V
31 2 V
31 3 V
32 2 V
31 3 V
31 2 V
32 3 V
31 2 V
31 3 V
32 2 V
31 3 V
31 2 V
32 3 V
31 2 V
31 3 V
31 3 V
32 2 V
31 3 V
31 3 V
32 2 V
31 3 V
31 3 V
32 2 V
31 3 V
31 3 V
32 3 V
31 3 V
31 2 V
32 3 V
31 3 V
31 3 V
31 3 V
32 3 V
31 2 V
31 3 V
32 3 V
31 3 V
31 3 V
32 3 V
31 3 V
31 3 V
32 3 V
31 3 V
31 3 V
32 3 V
31 3 V
31 3 V
31 3 V
32 3 V
31 3 V
31 3 V
32 3 V
31 4 V
31 3 V
32 3 V
31 3 V
31 3 V
32 3 V
31 4 V
31 3 V
32 3 V
31 3 V
31 3 V
31 4 V
32 3 V
31 3 V
31 4 V
32 3 V
31 3 V
31 3 V
32 4 V
31 3 V
0.500 UL
LT5
LTb
LT5
3087 403 M
263 0 V
350 1727 M
31 1 V
32 1 V
31 2 V
31 1 V
32 2 V
31 1 V
31 1 V
32 2 V
31 1 V
31 2 V
31 1 V
32 2 V
31 2 V
31 1 V
32 2 V
31 1 V
31 2 V
32 1 V
31 2 V
31 2 V
32 1 V
31 2 V
31 2 V
32 2 V
31 1 V
31 2 V
31 2 V
32 2 V
31 1 V
31 2 V
32 2 V
31 2 V
31 2 V
32 1 V
31 2 V
31 2 V
32 2 V
31 2 V
31 2 V
32 2 V
31 2 V
31 2 V
31 2 V
32 2 V
31 2 V
31 2 V
32 2 V
31 2 V
31 2 V
32 2 V
31 2 V
31 2 V
32 2 V
31 2 V
31 3 V
32 2 V
31 2 V
31 2 V
31 2 V
32 3 V
31 2 V
31 2 V
32 2 V
31 2 V
31 3 V
32 2 V
31 2 V
31 3 V
32 2 V
31 2 V
31 3 V
32 2 V
31 2 V
31 3 V
31 2 V
32 2 V
31 3 V
31 2 V
32 3 V
31 2 V
31 3 V
32 2 V
31 3 V
31 2 V
32 3 V
31 2 V
31 3 V
32 2 V
31 3 V
31 2 V
31 3 V
32 3 V
31 2 V
31 3 V
32 2 V
31 3 V
31 3 V
32 2 V
31 3 V
1.000 UL
LTb
350 300 M
3100 0 V
0 1760 V
-3100 0 V
350 300 L
1.000 UP
stroke
grestore
end
showpage
}}%
\put(3037,403){\makebox(0,0)[r]{\scriptsize$N_R=3, N_L=1, q=2, w=1$}}%
\put(3037,483){\makebox(0,0)[r]{\scriptsize$N_R=2, N_L=0, q=2, w=1$}}%
\put(3037,563){\makebox(0,0)[r]{\scriptsize$N_R=2, N_L=1, q=1, w=1$}}%
\put(3037,643){\makebox(0,0)[r]{\scriptsize$N_R=1, N_L=0, q=1, w=1$}}%
\put(3037,723){\makebox(0,0)[r]{\scriptsize$1^{st}$ e.s for $q=2, P=-$}}%
\put(3037,803){\makebox(0,0)[r]{\scriptsize$1^{st}$ e.s for $q=2, P=+$}}%
\put(3037,883){\makebox(0,0)[r]{\scriptsize g.s for $q=2, P=-$}}%
\put(3037,963){\makebox(0,0)[r]{\scriptsize g.s for $q=2, P=+$}}%
\put(3037,1043){\makebox(0,0)[r]{\scriptsize$1^{st}$ e.s for $q=1, P=-$}}%
\put(3037,1123){\makebox(0,0)[r]{\scriptsize g.s for $q=1, P=+$}}%
\put(3037,1203){\makebox(0,0)[r]{\scriptsize g.s for $q=1, P=-$}}%
\put(1900,50){\makebox(0,0){$l \sqrt{\sigma}$}}%
\put(100,1180){%
\special{ps: gsave currentpoint currentpoint translate
270 rotate neg exch neg exch translate}%
\makebox(0,0)[b]{\shortstack{$\sqrt{ E^2/ \sigma - (2 \pi q/ \sqrt{\sigma} l)^2}$}}%
\special{ps: currentpoint grestore moveto}%
}%
\put(3450,200){\makebox(0,0){ 4.5}}%
\put(2830,200){\makebox(0,0){ 4}}%
\put(2210,200){\makebox(0,0){ 3.5}}%
\put(1590,200){\makebox(0,0){ 3}}%
\put(970,200){\makebox(0,0){ 2.5}}%
\put(350,200){\makebox(0,0){ 2}}%
\put(300,2060){\makebox(0,0)[r]{ 9}}%
\put(300,1864){\makebox(0,0)[r]{ 8}}%
\put(300,1669){\makebox(0,0)[r]{ 7}}%
\put(300,1473){\makebox(0,0)[r]{ 6}}%
\put(300,1278){\makebox(0,0)[r]{ 5}}%
\put(300,1082){\makebox(0,0)[r]{ 4}}%
\put(300,887){\makebox(0,0)[r]{ 3}}%
\put(300,691){\makebox(0,0)[r]{ 2}}%
\put(300,496){\makebox(0,0)[r]{ 1}}%
\put(300,300){\makebox(0,0)[r]{ 0}}%
\end{picture}%
\endgroup
 